\documentclass[prb,twocolumn,showpacs,preprintnumbers,amsmath,amssymb]{revtex4}
\usepackage{graphicx}
\usepackage{dcolumn}
\usepackage{bm}

\begin{document}

\title{Electronic thermal transport in strongly correlated multilayered
nanostructures}

\author{J. K. Freericks}
\email{freericks@physics.georgetown.edu}
\homepage{http://www.physics.georgetown.edu/~jkf/index.html}
\affiliation{Department of Physics, Georgetown University,
Washington, DC 20057-0995}
\author{V. Zlati\'c}
\email{zlatic@ifs.hr}
\affiliation{Institute of Physics, Bijenicka c. 46, P. O. B. 304,
1000 Zagreb, Croatia}
\author{A. M. Shvaika}
\email{ashv@icmp.lviv.ua}
\homepage{http://ph.icmp.lviv.ua/~ashv/}
\affiliation{Institute for Condensed Matter Physics of the National
Academy of Sciences of Ukraine, 1 Svientsitskii Street, 79011 Lviv, Ukraine}

\begin{abstract}
The formalism for a linear-response
many-body treatment of the electronic contributions
to thermal transport is developed for multilayered nanostructures.  By
properly determining the local heat-current operator, it is possible to 
show that the Jonson-Mahan theorem for the bulk can be extended to 
inhomogeneous problems, so the various thermal-transport coefficient
integrands are related by powers of frequency (including all effects of
vertex corrections when appropriate).  We illustrate how to use this
formalism by showing how it applies to measurements of the Peltier effect,
the Seebeck effect, and the thermal conductance.
\end{abstract}


\maketitle

\section{Introduction}

As materials and device growth techniques mature and improve, it becomes
more possible to create artificial systems, composed of well-defined
numbers of flat planes of one material grown on top of another material.  
A device can be engineered by determining the different kinds of materials
to grow and the thicknesses of the different multilayers.
If we assume the growth process is perfect, so the planes are atomically
flat, with no interface roughness, then we have an inhomogeneous
quantum-mechanical problem to solve for the behavior of electrons in the system,
where the inhomogeneity lies in one dimension only.

The properties of devices grown in this fashion are more complicated if
some or all of the materials that make up the device are composed of
strongly correlated electron materials, where the properties of the
electrons cannot be described solely by an independent-particle picture like
band theory.  These systems are increasing in interest because bulk
strongly correlated materials exhibit exotic phenomena, and show promise
in demonstrating high tunability of their properties.  What is less studied
is how 
these properties can be modified by confinement/deconfinement effects that
are possible in multilayered nanostructures.

In addition, there has been little theoretical development of the thermal
transport effects in such multilayered systems.  Some evidence from 
examining the interface between two materials, indicates that thermal
transport effects in inhomogeneous systems can create large enhancements
to the performance\cite{rontani_sham}, but the theory has not been fully
developed within a Kubo-like context which allows the many-body aspects to be
treated fully.  Semiclassical approaches have also been 
employed\cite{mahan_nano}, but that theory is also inadequate to treat
strongly correlated materials.  Finally, the important
problem of phonon transport in
such systems has been examined extensively, but is beyond the scope
of what we cover in this work.

One of the most important results in bulk thermal transport is the 
Jonson-Mahan theorem\cite{jonson_mahan_1980,jonson_mahan_1990}, which
provides an exact relationship between different thermal transport 
coefficients.  Using the Jonson-Mahan theorem makes calculation of the 
thermal transport only slightly more complicated than the calculation of
the charge transport, and has enabled much of the theoretical work in
strongly correlated thermoelectrics.  Here we show how to generalize
the Jonson-Mahan theorem to multilayered
nanostructures, which also greatly simplifies
the calculation of the thermal transport coefficients.

The idea to use multilayered nanostructures, or more complicated geometries,
for enhancing thermoelectric performance of refrigerators or power
generators was proposed\cite{hicks_dresselhaus_1993} in the 1990s and
enhancements have been seen recently\cite{ventkatasubramanian}.  The focus
in that work was along the ideas of an electron-crystal-phonon-glass
approach, where the nanostructures are engineered to preserve the electronic
properties, while making the phonon transport similar to that in a disordered
glass.  It is possible that one can actually employ the nanostructure
engineering to produce {\it enhancements} to the electronic transport 
properties,
while simultaneously reducing the phonon thermal transport, so this basic 
approach may 
be pushed further than theorized in the original presentations.  One key
to being able to enhance the electronic properties, is to be able to tune
the electron correlation properties with a proper engineering of the
nanostructure and to engineer the charge redistribution at the
interfaces.  Before initiating such a program, one needs to be able
to properly calculate the thermal transport in a strongly correlated
device, and we derive the formalism for how to do this here.

The systems we describe in this work involve multilayered devices constructed
of atomically flat planes which can be composed of different materials.
Each system is inhomogeneous in the $z$-direction, which is the direction
where the planes are stacked. We take the left lead to be identical to the right
lead, so the system will have a ``mirror symmetry'', and the device will
have its chemical potential determined by that of the bulk leads.  We use
Roman letters ($i$, $j$, ...)
to denote the lattice sites within each plane ({\it i. e.}, the $x-y$ 
coordinates), and Greek letters ($\alpha$,
$\beta$, ...) to denote the individual planes ({\it i.e.}, the $z$-coordinate). 
We require the system to be
translationally invariant within each plane, and for the lattice structure
of each plane to be identical, so that the connection between planes is
between corresponding sites in the two planes, and is the same for each
site.  The latter requirement is by no means necessary, but it greatly 
simplifies the notation for the formalism, so we adopt it here.

We will consider three different types of Hamiltonians here: (i) the Hubbard
model\cite{hubbard_1963}; (ii) the Falicov-Kimball 
model\cite{falicov_kimball_1969};  and (iii) the periodic Anderson 
model\cite{anderson_1961}.  We use a multiple index $\alpha i$ to denote the
$i$th planar site on plane $\alpha$.  In the Hubbard model, we have 
conduction electrons, whose creation and annihilation operators are denoted
$c^\dagger_{\alpha i\sigma}$ and $c^{}_{\alpha i\sigma}$, respectively,
for electrons sitting at the lattice site denoted by $\alpha i$ and with
$z$-component of spin $\sigma$.  The Falicov-Kimball model has two kinds
of electrons: conduction electrons (which are described by similar operators
as in the Hubbard model, but without spin, for simplicity) and localized
electrons (also chosen to be spinless and created or destroyed by the
operators $f^\dagger_{\alpha i}$ and $f^{}_{\alpha i}$).  The periodic
Anderson model has spin-one-half conduction and $f$-electrons, which are denoted
by the familiar operators, except now all operators will have spin labels.
All models can be expressed as the sum of two terms in the Hamiltonian---an
inhomogeneous
hopping term and an interaction-hybridization term.  The inhomogeneous
hopping term is essentially the same for all three models.  It is 
\begin{eqnarray}
\mathcal{H}_{\rm hop}&=&-\sum_\alpha\sum_{i,j\in {\rm plane}}\sum_{\sigma}
t_{\alpha ij}^{\parallel}c^\dagger_{\alpha i\sigma}c^{}_{\alpha j\sigma}
\nonumber\\
&-&\sum_\alpha\sum_{i\in{\rm plane}}\sum_\sigma t^\perp_{\alpha\alpha+1}
c^\dagger_{\alpha i\sigma}c^{}_{\alpha+1 i\sigma}\nonumber\\
&-&\sum_\alpha\sum_{i\in{\rm plane}}\sum_\sigma t^\perp_{\alpha\alpha+1}
c^\dagger_{\alpha+1 i\sigma}c^{}_{\alpha i\sigma}
\label{eq: ham_hop}
\end{eqnarray}
where we do not include a sum over spin for the spinless Falicov-Kimball
model. We assume the hopping matrices are real symmetric matrices, and one
should note that the hopping between planes is only between neighboring planes
and between corresponding sites within the two planes.  The magnitudes of the
hopping matrices within the planes and between the planes can vary, but the
planar hopping matrices must be translationally invariant to go to a mixed
momentum-space--real-space basis, which is commonly done in these types
of problems.  If the planes are
square-lattice planes, then the underlying lattice topology will be that
of a simple cubic lattice, but the hopping need not be the same 
everywhere.

The interaction-hybridization term is different for each model.  For the
Hubbard model it is
\begin{equation}
\mathcal{H}^{\rm Hub}_{\rm int}=\sum_\alpha\sum_{i\in \rm plane}
U_\alpha c^\dagger_{\alpha i\uparrow}c^{}_{\alpha i\uparrow}
c^\dagger_{\alpha i\downarrow}c^{}_{\alpha i\downarrow},
\label{eq: hubbard_int}
\end{equation}
for the spinless Falicov-Kimball model it is
\begin{equation}
\mathcal{H}^{\rm FK}_{\rm int}=\sum_\alpha\sum_{i\in \rm plane}
U_\alpha c^\dagger_{\alpha i}c^{}_{\alpha i}
f^\dagger_{\alpha i}f^{}_{\alpha i},
\label{eq: fk_int}
\end{equation}
and for the periodic Anderson model it is
\begin{eqnarray}
\mathcal{H}^{\rm pam}_{\rm int}&=&\sum_\alpha\sum_{i\in \rm plane}\sum_\sigma
E_{F\alpha}f^\dagger_{\alpha i\sigma}f^{}_{\alpha i\sigma}\nonumber\\
&+&\sum_\alpha\sum_{i\in \rm plane}
U_\alpha f^\dagger_{\alpha i\uparrow}f^{}_{\alpha i\uparrow}
f^\dagger_{\alpha i\downarrow}f^{}_{\alpha i\downarrow}
\label{eq: pam_int}\\
&+&\sum_\alpha\sum_{i\in \rm plane}\sum_\sigma V^{\rm hyb}_{\alpha}
\left ( f^\dagger_{\alpha i\sigma}c^{}_{\alpha i\sigma}+
c^\dagger_{\alpha i\sigma}f^{}_{\alpha i\sigma}\right ).
\nonumber
\end{eqnarray}
For the Falicov-Kimball model, we often replace the term $f^\dagger_{\alpha i}
f^{}_{\alpha i}$ by the symbol $w_{\alpha i}$ which equals 0 if no localized
electrons are at site $\alpha i$ and equals 1 if a localized electron is
at site $\alpha i$.
All interaction and hybridization terms can vary from plane to plane, but
they must be the same for every lattice site within the planes to preserve
translational invariance within the planes.  The total Hamiltonian is then
\begin{equation}
\mathcal{H}=\mathcal{H}_{\rm hop}+\mathcal{H}_{\rm int}-\mu\mathcal{N},
\label{eq: ham_total}
\end{equation}
for all of the models.  The symbol $\mu$ is the chemical potential, and
$\mathcal{N}$ denotes the electron number operator, chosen to be the 
total conduction electron number operator for the Hubbard and Falicov-Kimball
models and the total electron number operator for the periodic Anderson
model (we work in a canonical ensemble for the $f$-electrons in the
Falicov-Kimball model, so no site-energy or chemical potential is needed
for those particles).

In Section II, we present a description of electronic charge reconstruction,
which naturally occurs in any multilayered device that can be grown for
thermoelectric properties.  This section briefly reviews the current status of 
such calculations, and describes their impact on the thermal transport; in
particular, it fixes the notation for the internal electrostatic potentials
associated with the electronic charge reconstruction.
Section III provides the main arguments for developing the multilayered
generalization of the Jonson-Mahan theorem.  Section IV applies the formalism
to three classic experiments---the Peltier effect, the Seebeck effect, and the
thermal conductivity.  Section V presents our conclusions and describes areas
for further work.

\section{Electronic charge reconstruction in multilayered nanostructures}

The Schottky effect~\cite{schottky_1940}, is a well-known effect in the
semiconductor community, where charge is redistributed between a 
semiconductor and a metal at a semiconductor-metal interface due
to a bulk chemical potential mismatch between the two materials.  The
charge rearrangement creates a screened dipole layer at the interface
resulting in a final state with a static inhomogeneous redistribution of 
charge through the system.  Recently, the phenomenon has been revisited
in the context of strongly correlated materials~\cite{freericks_sinis},
where it has been called electronic charge 
reconstruction~\cite{millis_okamoto,macdonald}.  If we imagine a multilayered 
nanostructure, composed of metallic leads sandwiching a barrier region,
which is a strongly correlated material, then the chemical potential
of the device is fixed by the chemical potential of the leads.  If the
chemical potential of the barrier is different, then the system must
undergo an electronic charge reconstruction.  In particular, since
the temperature dependence of the chemical potential should be different
in the two different materials, even if the chemical potentials match
at one temperature, they will not match at other temperatures, and a charge
redistribution will take place.

In this work, we will focus on problems that have ``mirror symmetry'' for
the leads, so the
lead to the left is made of the same material as the lead to 
the right.  In this case, the total Coulomb
potential energy, due to all electric
fields, goes to zero when one is far from the interfaces, because all
of the charge rearrangement is localized at the interfaces, and the
whole system is charge neutral.  If we were to examine systems with
different materials for the left and right leads, then the electrochemical
potential of the system will be the average of the left and right bulk
chemical potentials,
which creates some additional complications, but does not change the
basic strategies or formulas, although, the Seebeck effect needs to be
defined and analyzed with care\cite{mahan_seebeck}.

The approach to describe the electronic charge reconstruction is
a semiclassical one.  We solve the problem for local electron interactions
exactly, but treat the long-range Coulomb interaction in a mean-field
fashion.  The strategy we use is to first calculate the
electronic charge on each plane via an inhomogeneous Green's function approach 
(in dynamical mean-field theory, the technique of
Potthoff and Nolting~\cite{potthoff_nolting_1999} is used).  If possible,
one uses a Matsubara-frequency approach, because the numerics are usually under
better control than real-axis approaches, but this is just a matter of
convenience, not necessity.  Next, we find the charge deviation
on each plane; namely, we determine whether extra charge has entered or
left the plane.  Since the positive background charge of the ions remains
the same, the charge deviation will give rise to an electric field.  There
are two different ways to treat this field.  The simplest is to assume the
electric charge is uniformly spread over the
plane~\cite{freericks_sinis}.  Then the electric field is
constant, perpendicular to the plane, and pointing away from it
in both directions if the net charge density is positive, while pointing
toward the plane if the net charge density is negative. The second method
uses the actual distribution of the ions, and the spatial profile of the
electrons, if available, to calculate the charge~\cite{millis_okamoto}.  
This approach is closer to an Ewald-like summation~\cite{ewald}
of the charge densities.  The two treatments should yield similar results.

\begin{figure}[th]
\centerline{\includegraphics[width=3.0in,angle=0]{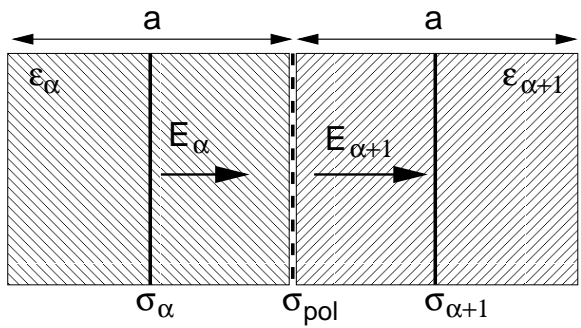}}
\vspace*{8pt}
\caption{Geometry taken for the classical electrostatics problem.
We show the blow up of two planes, $\alpha$ and $\alpha+1$.  Assuming the
net surface 
charge density on plane $\alpha$ is $(\rho_{\alpha}-\rho_{\alpha}^{\rm bulk})a
=\sigma_\alpha$ (located along the plane running through position $\alpha$)
and the relative permittivity is $\epsilon_{\rm r \alpha}$ (and similarly for
the $\alpha+1$ plane), then the change in polarization at the interface between
the two dielectric planes induces a polarization
charge on the interface (denoted $\sigma_{\rm pol}$) that leads to a
discontinuous jump in the electric field halfway between the two lattice
planes (at the position $\alpha+1/2$).  
Once the fields are known, we integrate to get the electric potentials.
Note the discontinuity in the electric field occurs at the {\it midpoint}
between the two lattice planes.
\label{fig: pol_charge}}
\end{figure}

In this work, we will choose the ``constant plane of charge'' description for
determining the electric fields.  This allows us to determine
simple analytic expressions for the electric fields---for example, the
magnitude of the constant field, emanating
from the $\alpha$th plane of charge is
\begin{equation}
|{\bf E}_\alpha|=\frac{|e||\rho_\alpha-\rho^{\rm bulk}_\alpha|a}{2\epsilon_0
\epsilon_{\rm r\alpha}},
\label{eq: electric_field_planes}
\end{equation}
where $e<0$ is the charge of the electron,
$\rho_\alpha$ is the quantum-mechanically calculated electron number
density at plane $\alpha$, $\rho_\alpha^{\rm bulk}$ is the bulk electron
number density for the material that plane $\alpha$ is composed of
(equal to the positive background charge on the plane),
$\epsilon_0$ is the permittivity of free space, and $\epsilon_{\rm r\alpha}$
is the relative permittivity of plane $\alpha$.
The contribution to the electric potential $V^{\rm c}(z)$  from this field
satisfies
\begin{equation}
E=-\frac{d}{dz}V^{\rm c}(z).
\label{eq: potential_diffeq}
\end{equation}
Since the electric field is constant in magnitude, it is straightforward
to compute the contribution to the Coulomb potential at plane $\beta$
due to the change in the charge density at plane $\alpha$ (but one needs to keep
track of the signs of the fields or equivalently
the relative order of $\alpha$ with respect to $\beta$):
\begin{eqnarray}
V^{\rm c}_{\beta}(\alpha)&=&\frac{|e|(\rho_\alpha-\rho_\alpha^{\rm bulk})
a}
{2\epsilon_0}\nonumber\\
&\times&\left \{\begin{array}{l l}
\sum_{\gamma=\alpha+1}^\beta [\frac{1}{2\epsilon_{\rm r \gamma}}
+\frac{1}{2\epsilon_{\rm r \gamma-1}}],&
\quad\beta>\alpha\\
0,&\quad\beta=\alpha\\
\sum_{\gamma=\alpha-1}^\beta [\frac{1}{2\epsilon_{\rm r \gamma}}
+\frac{1}{2\epsilon_{\rm r \gamma+1}}],&
\quad\beta<\alpha
\end{array}\right. .
\label{eq: potential_ab}
\end{eqnarray}
Note that if the relative permittivity $\epsilon_r$
is a constant, independent of the
planes, then the potential energy is a linear function of the $z$-coordinate,
proportional to
$-|{\bf z}_\alpha-{\bf z}_\beta|/\epsilon_{\rm r}$ as one might expect.
The reason why we need to sum over two terms in the summands in
Eq.~(\ref{eq: potential_ab}) is because we envision the $\alpha$th
plane of charge to be infinitesimally thick,
and go through the lattice sites of plane $\alpha$, but we assume
the dielectric has
a thickness of $a$ and is centered around each plane of atoms.
Hence, if the permittivity changes from one plane to another, a polarization
charge develops halfway between the two planes where the dielectric is
changing, and the electric field has a discontinuity at that point
({\it i. e.}, at the position $\alpha+1/2$, see Fig.~\ref{fig: pol_charge}).

It is actually the potential energy $-|e|V^{\rm c}_\alpha=V_\alpha$ that 
shifts the chemical potential at each planar site. We define a parameter
\begin{equation}
e_{\rm Schot}(\alpha)=\frac{e^2 a}{2\epsilon_0\epsilon_{\rm r\alpha}},
\label{eq: eschot_def}
\end{equation}
which controls how the extra charge density decays away from the interfaces.
The parameter $e_{\rm Schot}$ has the units of an energy multiplied by an
area; the product of $e_{\rm Schot}$ with the local density of states has units 
of the inverse of a length, and this is what determines the decay length of
the charge profile.
Using this parameter, we can immediately calculate the
potential energy due to the Coulomb interaction (evaluated in a mean-field
fashion)
\begin{eqnarray}
V_{\beta}&=&-\sum_\alpha(\rho_\alpha-\rho_\alpha^{\rm bulk})
\label{eq: potential_energy_schottky}\\
&\times&
\left \{\begin{array}{l l}
\sum_{\gamma=\alpha+1}^\beta \frac{1}{2}
[e_{\rm Schot}(\gamma)+e_{\rm Schot}(\gamma-1)],& \quad\beta>\alpha\\
0,&\quad\beta=\alpha\\
\sum_{\gamma=\alpha-1}^\beta \frac{1}{2}
[e_{\rm Schot}(\gamma)+e_{\rm Schot}(\gamma+1)], & \quad\beta<\alpha
\end{array}\right. .
\nonumber
\end{eqnarray}
Note that a similar analysis can be carried out if one uses the Ewald-like
technique for determining the charge reconstruction. 

These potential energies modify the Hamiltonian by the long-range Coulomb
interaction of the charge reconstruction.  The additional piece of the
Hamiltonian  (due to the charge rearrangement) is
\begin{equation}
\mathcal{H}_{\rm charge}=\sum_\alpha V_\alpha \sum_{i\in {\rm plane}}
c^\dagger_{\alpha i}c^{}_{\alpha i}.
\label{eq: charge_ham}
\end{equation}
Hence, they can be treated by shifting the chemical potential $\mu\rightarrow
\mu-V_\alpha$ on each plane depending on what the Coulomb potential energy
is for the given plane. For consistency, we must have that the potentials
go to zero as we move far enough into either of the leads (for the
mirror-symmetric case). This requirement
enforces overall charge conservation---any charge that moves out of the
barrier remains in the leads, localized close to the interface, and
{\it vice versa}. Of course, the potentials $V_\alpha$ that appear in the
electronic charge reconstruction Hamiltonian in Eq.~(\ref{eq: charge_ham})
must be determined self-consistently.  Achieving this goal requires care
in setting up the iterative algorithm.

There will be no electronic charge reconstruction if the chemical potentials
in the bulk of both the leads and the barrier match.  In order to have
freedom to adjust the mismatch of the chemical potentials, we need to
be able to change the value of the band zero of the barrier region
relative to the band zero of the leads.  This parameter is called
$\Delta E_{F\alpha}$, which vanishes
in the leads, and is generically a nonzero constant in the barrier (independent
of the temperature or the charge rearrangement). Hence we add an additional
term 
\begin{equation}
\mathcal{H}_{\rm offset}=-\sum_\alpha\sum_{i\in \rm plane}
\Delta E_{F\alpha}c^\dagger_{\alpha i}c^{}_{\alpha i}
\label{eq: h_offset}
\end{equation}
to the Hamiltonian.
Although this term appears similar to the Coulomb potential term
in Eq.~(\ref{eq: charge_ham}), the key observation
is that this term is fixed and does not change with any parameters of the
system, whereas the plane potentials $V_\alpha$ need to be readjusted as
the parameters change, to achieve a self-consistent solution of the problem.
Note that we set $\Delta E_{F\alpha}=0$ in the leads to the right and to the 
left.

\begin{figure}[th]
\centerline{\includegraphics[width=3.5in,angle=0]{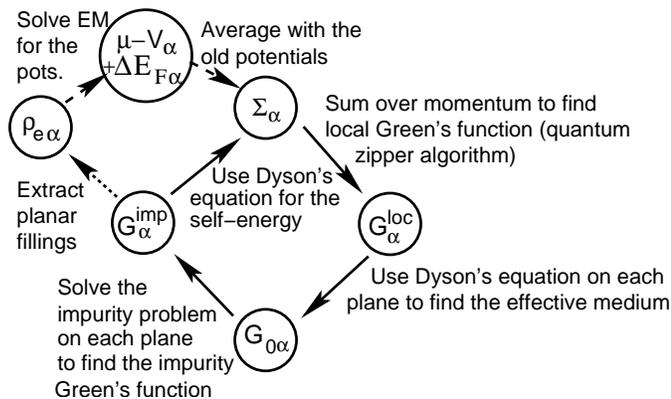}}
\vspace*{8pt}
\caption{Flow diagram for the DMFT algorithm in a multilayered
nanostructure with electronic charge reconstruction.  We determine the
charges on each plane, determine how they differ from the bulk charge on
the plane to find the excess or deficit charge.  Then we use classical
electromagnetism to find the electric potentials on each plane and finally
the contribution of the potential energy to the electrochemical potential
on each plane. Then we average the potentials with a large damping factor
so that the potentials are updated slowly.  This is then input into the
next loop of the main DMFT algorithm, which is unchanged from cases where there
is no electronic charge reconstruction.
\label{fig: dmft_algorithm3}}
\end{figure}

In this contribution, we will not discuss how to actually solve for the 
electronic charge reconstruction in detail.  One can imagine a number of 
different approaches to this problem, ranging from direct means of solving
the quantum-mechanical problem on finite-sized stacked planes, to other
techniques like the inhomogeneous
dynamical mean-field theory approach.  The DMFT approach
has been quite successful in examining these kinds of problems, and the
formalism only requires that the self-energy remain local (although it can
vary from plane to plane).  Then, by performing a Fourier transform to
momentum space for the planar coordinates, one decouples the planar motion
from the longitudinal motion.  Hence, the problem reduces to a series of
quasi-one-dimensional inhomogeneous problems, which can be solved by
using the renormalized perturbation expansion\cite{rpe} (sometimes called the
quantum zipper algorithm\cite{freericks_2004}). Details for such an approach
have already appeared\cite{freericks_sinis,freericks_2004} and are briefly 
reviewed below.

The DMFT algorithm is given in Fig.~\ref{fig: dmft_algorithm3}.
If there is a separate algorithm available
for the Matsubara frequency Green's functions, then the upper left loop (which
determines the electronic charge reconstruction) is used
on the imaginary axis, and we don't need it on the real axis.  If such an
algorithm does not exist (such as when calculations are performed with the
numerical renormalization group), then we would use the entire loop on the 
real axis. The new steps to find the electronic charge reconstruction are to
first find the electron density on each plane.  Then we subtract the bulk
charge density of each plane to find the excess or deficit charge on the
given plane.  Once the change in charge density is known, we can calculate
the electrical potential, and then the contribution to the potential energy.
This gets added to the chemical potential to determine the electrochemical
potential at each plane.  

When numerical results are generated~\cite{freericks_sinis,chen_freericks}
(see those references for numerical issues in the algorithm),
we find that usually the electronic charge reconstruction does not change
significantly at low temperature, and that the size of the charge
deviation grows as the mismatch between the chemical potentials grows (governed
in part by the size of $\Delta E_{F\alpha}$ in the barrier).  Such results 
are similar to
what one would expect, but most of the calculations have taken place
in systems where the charge density is not too sensitive to changes in
the chemical potential.  The effects may be different in systems with 
either Mott insulators, or doped Mott insulators, which can be brought
close to the insulating phase via the electronic charge reconstruction.
In addition, electron-phonon coupled systems can develop a strong sensitivity
of the charge to the chemical potential when the coupling is large, which
may be an interesting case to examine as well.

Ultimately, we are most interested in the transport of charge and heat
through the device.
In order to calculate the transport, we need to evaluate
the real-axis results for the self-energies and Green's functions
of a nanostructure with an electronic charge
reconstruction.  Unfortunately, the algorithms used when there is no electronic
charge reconstruction~\cite{potthoff_nolting_1999,freericks_2004} cannot be 
simply employed for this case.  The reason why
is that the presence of the different potentials $V_\alpha$ on each plane
causes the nature of the integrands over the two-dimensional density of states
to have a different singular behavior than they had before.  In a system without
electronic charge reconstruction, the singularities in the integrand could be 
square-root-like, which are removed by a simple
variable change using trigonometric or hyperbolic functions.  Now, the 
singularities are poles (because the denominators
are shifted by the potentials at a given plane, so they vanish at
different energies, and give rise to a different
singular behavior), and we need to evaluate
all integrals in a principal-value sense, where the real part is integrated
with a symmetric grid around each pole, and the imaginary part has a
delta-function contribution that needs to be included.  This is challenging
to implement numerically, because the locations of the poles are different
on different planes, and can vary from one iteration to the next.  Details
for how to deal with such a sophistication will appear elsewhere, since they
are beyond the scope of this work.

\section{Proof of the inhomogeneous Jonson-Mahan theorem}

It is important to examine how the linear-response transport 
formalism\cite{freericks_2004} is modified by the presence of an
electronic charge reconstruction.  We have taken the
chemical potential as a constant throughout
the multilayered nanostructure for thermodynamic equilibrium. One can directly
show that the device carries no longitudinal
charge current even though there are nonzero
electric fields arising from the electronic charge reconstruction (see
Appendix A for a proof when the self-energy
is local). No current flows because the putative
current driven by the internal electric fields is canceled by an equal magnitude
but oppositely directed current driven by the concentration gradients.  The
standard way to describe this result is via a phenomenological equation
(for the case with no thermal gradients)~\cite{onsager_1931}
\begin{eqnarray}
\langle j^{\rm c}\rangle&=&a\sum_\beta\sigma_{\alpha\beta}E_\beta
-a|e|\sum_\beta\mathcal{D}_{\alpha\beta}\frac{\rho_{\beta+1}-\rho_\beta}{a}
\nonumber\\
&=&-\frac{a}{|e|}
\sum_\beta\sigma_{\alpha\beta}\frac{\tilde\mu_{\beta+1}-\tilde\mu_\beta}
{a},
\label{eq: onsager_phenom}
\end{eqnarray}
where $\mathcal{D}_{\alpha\beta}$ is the diffusion constant for Fick's law
of diffusion~\cite{fick_1855},
and the second equality follows from the
Einstein relation~\cite{einstein_1905}
(or more correctly the
Nernst-Einstein-Smoluchowski relation~\cite{nernst_1889,smoluchowski_1906})
which relates the diffusion constant to the conductivity via
\begin{equation}
\sigma_{\alpha\beta}=e^2\mathcal{D}_{\alpha\beta}{d\rho_\beta/d\mu};
\label{eq: einstein_relat}
\end{equation}
both quantities are matrices, with indices given by the planes of the
multilayered nanostructure.
The symbol $\tilde\mu_\alpha=\mu-V_\alpha$ is called the electrochemical
potential.
The Einstein relation can be derived by relating the gradient
with respect to the chemical potential to the gradient with respect to the
number concentration via the chain rule: $d\mu/dz=(d\mu/d\rho)d\rho/dz$, and the
fact that the current vanishes in equilibrium~\cite{luttinger_1964}.

Eq.~(\ref{eq: onsager_phenom}) implies that the condition for there to be
no charge current is simply $d \tilde\mu/dz=0$. The chemical potential is
a constant, but it does vary with the filling, so if there is a change in
electron concentration, then $d\tilde\mu/dz=(d\mu/d\rho)d\rho/dz-dV(z)/dz$,
so the force from the electric field will be balanced by the force from
the change in electron concentration. In addition,
note that the current vanishes
no matter how large the variation in the concentration is ({\it i. e.},
beyond the linear-response regime), so the conclusion is that the current
generated by the internal electric field is always canceled by the current
generated by the change in the electron concentration.  Hence, for a
linear-response treatment of transport, we can ignore the forces due to the
internal electric fields and the concentration gradients, because they always
cancel, and we can limit our focus to the effects of the external electric field
only.  This then implies that all of the analysis performed previously for
the charge current\cite{freericks_2004} continues to be valid, and because the 
form of the charge
current is unchanged when we have electronic charge reconstruction, the
Kubo formula is identical as it was before (with the effects of the potentials
$V_\alpha$ included, of course).

The basic observation needed for a thermoelectric device is that
there is a difference between the weighting factors that determine the
bulk charge current and heat current. The charge current is weighted by the
electron velocity, while the heat current is weighted by the velocity
multiplied by the kinetic energy minus the chemical potential plus
a term from the potential energy. Hence, one
can create charge current without heat current, or {\it vice versa}; by
carefully engineering the way electrons move through the device, one can
control both the energy and charge flow, which is useful for different 
types of applications like refrigeration or power generation. A typical device
has two legs, one using electrons as the charge carriers and one using holes
as the charge carriers.  Current flows through the device in a loop, but the 
net heat flows in one direction only, which allows the device to function.

In this contribution, we concentrate on multilayered nanostructures, which can 
be used to compose one of the legs of the thermoelectric device.
We also concentrate
solely on electronic transport mechanisms.  In most thermoelectrics,
the thermal conductivity from phonons can be large enough to significantly
reduce the figure-of-merit. 
It is expected that the phonon thermal conductivity will be further reduced
in a nanostructure, because the interfaces in the nanostructures will
cause significant phonon scattering if the masses of the ions in the different
materials have a large mismatch~\cite{hicks_dresselhaus_1993}, but we do not
discuss this issue further.

There is no simple way to derive the response of a strongly correlated
system to both electrical fields and thermal gradients.  The reason
why is that the thermal gradient cannot be added as a field to the
Hamiltonian like the electric field can, hence there is no way to
follow the simple Kubo response theory developed for the charge current
in an electric field (because the linear-response approach evaluates
correlation functions at a fixed temperature, and a variation of the temperature
with position is problematic to include within the formalism).  Luttinger
sorted out a reasonable plan of action for how one can nevertheless
proceed\cite{luttinger_1964}. We couple a fictitious field to
the heat-current operator, analogous to the vector potential that couples to
the charge current operator, and determine the linear response with respect to
both fields.  Then, we compare the Kubo response to a phenomenological
set of equations that relate the charge and heat currents to the electric
field and the gradient of the temperature.  We then identify the
relevant transport coefficients and how they are expressed in terms of
correlation functions. 

In multilayered nanostructures, there always is an electronic charge 
reconstruction, because the bulk chemical potentials for the leads and the
barrier will have different $T$ dependence, and hence cannot always be equal
(the only exception is for particle-hole symmetry at half-filling, but
there the thermopower vanishes, so that case is uninteresting for thermal
transport). Hence the Hamiltonian must be modified to include the potential
energy $V_\alpha$ on each plane, and the band offsets $\Delta E_F$, as described
in Sec.~II [{\it i. e.}, we add $\sum_{\alpha i}(V_\alpha-\Delta E_{F\alpha})
c^\dagger_{\alpha i}c^{}_{\alpha i}$ to $\mathcal{H}$]. The band offsets are
independent of $T$, and represent the difference in the band zeroes for the
leads and the material placed at plane $\alpha$.  The potential energies
$V_{\alpha}$ do depend on $T$, but they do not create any currents, because they
correspond to the static potential associated with the electronic charge
reconstruction (and the diffusion current generated by the change in electron
concentration cancels the current from the internal electric field; see Appendix
A). But the Coulomb potentials do create
internal electric fields that maintain the electronic
charge redistribution amongst the planes.

The phenomenological study of currents caused by external
electric fields or temperature
gradients has been examined since the early 1800s. It was found that
an electric field can drive a charge current (which is essentially Ohm's
law~\cite{ohm_1827}
with the conductivity as the phenomenological constant) and it can drive a
heat current because the electrons carry heat with
them as they move through the material (this phenomenon is called the
Peltier effect~\cite{peltier_1834}).
Similarly, a temperature gradient can drive heat
conduction with the phenomenological thermal conductivity (called Fourier's
law~\cite{fourier_1822}), and because
the electronic contribution to the
heat current generically carries charge, a temperature gradient can
generate a charge current (called the Seebeck effect~\cite{seebeck_1823}).
The phenomenological equations for the (linear response)
longitudinal transport in a multilayered nanostructure are then
($j_\alpha^{}$ is the longitudinal number current, $j_\alpha^{\rm c}$ is the 
longitudinal charge current, and $j^{\rm Q}_\alpha$ is the longitudinal
heat current)
\begin{eqnarray}
\langle j_\alpha^{\rm c}\rangle&=&-|e|\langle
j_\alpha\rangle\nonumber\\
&=&
|e|a\sum_\beta L_{11\alpha\beta}\left [ \frac{d\mu_\beta}{dT}
\frac{T_{\beta+1}-T_\beta}{a}+|e|E_\beta\right ]\nonumber\\
&+&|e|a\sum_\beta L_{12\alpha\beta}
\frac{T_{\beta+1}-T_\beta}{aT_\beta},
\label{eq: charge_phenom}\\
\langle j_\alpha^{\rm Q}\rangle&=&
-a\sum_\beta L_{21\alpha\beta}\left [ \frac{d\mu_\beta}{dT}
\frac{T_{\beta+1}-T_\beta}{a}+|e|E_\beta\right ]\nonumber\\
&-&a\sum_\beta L_{22\alpha\beta}
\frac{T_{\beta+1}-T_\beta}{aT_\beta},\nonumber\\
\label{eq: heat_phenom}
\end{eqnarray}
where the indices $\alpha$ and $\beta$ denote the planar sites (or the midpoint
between planar sites, as clarified below), the term 
$(T_{\beta+1}-T_\beta)/a$ is the
discretized approximation to the temperature gradient and the $L_{ij}$
coefficients can be thought of as the phenomenological parameters.
We define the symbol $\mu_\beta=\mu-V_\beta+\Delta E_{F\beta}$, which may be
thought of as the ``local chemical potential'' for plane $\beta$. 
The origin of the temperature derivative of $\mu_\beta$ entering into the
phenomenological equations arises from the conventional $\nabla\mu$ term, which
becomes $\nabla Td\mu/dT$ when the system is placed in a thermal gradient. The
spatial derivative of the $V_\beta$ terms does not drive any current, because it
cancels with the current driven by the equilibrium concentration gradient
(which we did not include in the above phenomenological equations),
so the electric field $E_\beta$ is the {\it external field} applied
to the device (this is valid only in the linear-response regime of a small
external electric field).
Note, that there is a simple way to understand the signs that appear in
Eqs.~(\ref{eq: charge_phenom}) and (\ref{eq: heat_phenom}).  First consider
the external electric field, which can be written as the negative gradient of
the electric potential.  The current (whether of electrons or of holes), always
runs down the potential hill.  Since the conductivity is always positive,
the first term in Eq.~(\ref{eq: charge_phenom}) must have a positive sign.
The thermoelectric number current also runs downhill, so it is proportional
to the negative temperature gradient.  For electrons, the charge current
is $-|e|$ times the number current, which gives rise to the positive
sign for the last term in Eq.~(\ref{eq: charge_phenom}). Similarly,
the thermal conductivity runs down the temperature ``hill'', so it has
a negative sign in front of it.  The Peltier effect term is the hardest
to understand, but because the electrons are negatively charged, they
actually move up the potential hill (the charge current runs down the
hill because the electrons are negatively charged), so the heat is
carried up the hill, and hence there is a minus sign in front of the term
(recall the electric field is the negative gradient of the potential).

Our next step is to determine how to represent the thermal transport
coefficients $L_{ij}$ in terms of many-body correlation functions.
This has already been done for the first coefficient\cite{freericks_2004}, 
which is proportional
to the conductivity matrix, and is represented by a current-current
correlation function: $\sigma_{\alpha\beta}=e^2L_{11\alpha\beta}$
(the modification of the Hamiltonian by the electronic charge
reconstruction has no effect on the form of the charge current, or on the
form of the correlations functions, but obviously creates additional
scattering).
Since this coefficient arises from an electric field, which can be
added to the Hamiltonian, the derivation is rigorous.  Similarly, if
we follow all the steps in Ref.~\onlinecite{freericks_2004}
that led up to the derivation of the conductivity
matrix, but we examined the expectation value of the heat-current operator
instead of the charge-current operator, we would find that the $L_{21}$
correlation function was identical to the $L_{11}$ correlation function
except that it is a heat-current--charge-current correlation function
instead of a charge-current--charge-current correlation function.

As we discussed above, there is no complete theory to determine the
$L_{12}$ and $L_{22}$ coefficients for the phenomenological transport
equations. But, classical nonequilibrium statistical mechanics has
proved that there is a reciprocal relation between the ``cross'' terms in
the transport equations~\cite{onsager_1931}.  Written
in the form we have them, this 
relation says that $L_{21}=L_{12}$. Knowing the form for $L_{21}$, we
then conclude that $L_{12}$ is the charge-current--heat-current correlation
function.  Keeping within this same vein, the natural conclusion is that the
final transport coefficient $L_{22}$ is a heat-current--heat-current correlation
function (but there is no rigorous derivation of this result).

In order to derive the local charge and heat current operators, we must 
formulate the transport problem in real space. Unlike the bulk case,
where the procedure is completely well-defined, there are a number of
different possible ways to try to derive the local current operators.
The bulk number current operator is found by taking the commutator of
the number polarization operator with the Hamiltonian; this guarantees that
the equation of continuity holds, and it also implies that the number current
is conserved through the system.  The number polarization operator is
\begin{equation}
\Pi^{\rm number}=\sum_{\alpha i}{\bf R}_{\alpha i}(c^\dagger_{\alpha i}
c^{}_{\alpha i}+f^\dagger_{\alpha i}f^{}_{\alpha i}),
\label{eq: number_pol}
\end{equation}
where we dropped the spin index for simplicity, and where ${\bf R}_{\alpha i}$
is the position vector of the site labeled by $\alpha i$. The $f^\dagger f$ term
enters for the Falicov-Kimball and periodic Anderson models, but not for the
Hubbard model. Since the number polarization operator depends only on the 
number operators, it commutes with all number operators in the interaction 
Hamiltonian.  It turns out that the form in Eq.~(\ref{eq: number_pol}) also
commutes with the hybridization term in the periodic Anderson model
because the hybridization is on-site
only.  Hence, the bulk number current operator is the same for all
models we examine here, and arises solely from the commutator of the 
$z$-component of the polarization operator
with the hopping Hamiltonian.  Performing the commutator is straightforward, and
leads to ${j}=\sum_\alpha{j}_\alpha,$ with
\begin{equation}
{j}_\alpha=-iat^\perp_{\alpha\alpha+1}\sum_{i\in {\rm plane}}
\left ( c^\dagger_{\alpha i}c^{}_{\alpha+1 i}-c^\dagger_{\alpha+1 i}
c^{}_{\alpha i}\right ).
\label{eq: current_long}
\end{equation}
Note that the subscript $\alpha$ on the
current operator denotes the total current operator flowing through the
$\alpha$th plane, and does not indicate a Cartesian coordinate of the
current operator; the current operator is always taken in the $z$-direction
for the longitudinal flow. The current at plane $\alpha$ is thus defined
to be the total number of electrons flowing to the left minus the total
flowing to the right (here the current operator at plane $\alpha$ is determined
by the number of electrons flowing to the right or to the left through
the $\alpha$th and $\alpha+1$st planes). 

A comment is in order about the choice given in Eq.~(\ref{eq: current_long})
for the current associated with the $\alpha$th plane.  Note that the
form chosen is not the same as the choice that would arise from taking the
commutator of the ``local'' polarization operator (at the $\alpha$th plane) with
the Hamiltonian.  The direct result from the commutator $\hat j_\alpha=i
[\mathcal{H},\sum_{i\in \rm plane}z_\alpha (c^\dagger_{\alpha i}c^{}_{\alpha i}
+f^\dagger_{\alpha i}f^{}_{\alpha i})]$
\begin{eqnarray}
\hat j_{\alpha}&=&-i\sum_{i\in \rm plane}
[t^\perp_{\alpha\alpha+1}(c^\dagger_{\alpha+1i} c^{}_{\alpha i}
-c^\dagger_{\alpha i} c^{}_{\alpha+1i})\nonumber\\
&+&
t^\perp_{\alpha-1\alpha}(c^\dagger_{\alpha-1i} c^{}_{\alpha i}-
c^\dagger_{\alpha i} c^{}_{\alpha-1 i})]z_{\alpha},
\label{eq: current_loc_alpha}
\end{eqnarray}
does not seem reasonable,
because it is weighted by the $z$-coordinate of the $\alpha$th plane,
rather than involving the difference of currents moving in opposite
directions (at the $\alpha$th plane).  When we have full translational
symmetry, we derive the
conventional form for the current operator by shifting the spatial index
of one of the terms, to explicitly carry out the cancellation of the
spatial coordinates (just take the summation of the above result over
$\alpha$, and shift $\alpha\rightarrow\alpha+1$ in the last two terms).
More reflection on this issue, shows that the explicit form of the local
current operator that enters the Kubo formula actually originates from the
coupling term $-{\bf j}\cdot {\bf A}$ term that corresponds to the 
perturbation of the
Hamiltonian due to the electric field in a gauge where the scalar potential
vanishes; this is because we evaluate the
expectation value of the total current with the perturbation of the
Hamiltonian due to the external field and that field enters via the vector
potential value at a specific plane.  Hence the conductivity matrix is defined
from the piece of the total current operator that couples to the field
at plane $\alpha$ and, since the total current will be the sum of the currents
at each plane, the current-current correlation function for
the conductivity matrix involves the
local current operators that couple to the vector potential. Thus,
we choose the perturbation of the Hamiltonian to be
\begin{equation}
\mathcal{H}^\prime(t)=-i|e|a\sum_{\alpha i} t^\perp_{\alpha\alpha+1}
(c^\dagger_{\alpha+1 i} c^{}_{\alpha i}-
c^\dagger_{\alpha i} c^{}_{\alpha+1i})A_\alpha(t),
\label{eq: current_pert_local}
\end{equation}
where we have taken the vector potential along the $z$-direction, and
independent of the intraplane coordinates, because the field is uniform for each
plane. We feel this choice makes good physical sense because we couple the
vector potential to the physical current between the $\alpha$th and
$\alpha+1$st planes. Alternatively, one can view this as a coupling of the
current between the $\alpha$th and $\alpha+1$st plane to the electric vector
potential located halfway
between those two planes (in this interpretation, we would use
$[A_\alpha+A_{\alpha+1}]/2$ as the coupling field).
Finally, one can take a symmetrized version of the
local current operator to be 
\begin{eqnarray}
j^{\rm sym}_\alpha
&=&-iat^\perp_{\alpha-1\alpha}\sum_{i\in\rm plane}
(c^\dagger_{\alpha i}c^{}_{\alpha-1 i}-c^\dagger_{\alpha-1 i}c^{}_{\alpha i})/2
\nonumber\\
&-&iat^\perp_{\alpha\alpha+1}\sum_{i\in\rm plane}
(c^\dagger_{\alpha+1 i}c^{}_{\alpha i}-c^\dagger_{\alpha i}c^{}_{\alpha+1 i})/2,
\label{eq: j_sym}
\end{eqnarray}
corresponding to the average of the currents located just to the left and to
the right of
plane $\alpha$.  This choice sounds like the most physical choice, but the
calculations for it are somewhat more complicated, and it is not likely the
end results are too different from our first choice.  The difference
between the two choices is actually quite simple.  In the first
approach, one should envision the spatial indices $\alpha$ and $\beta$ to
correspond to $z_\alpha+a/2$ and $z_\beta+a/2$; that is, they are
shifted to the right by half the distance between the planes.  In the
second, symmetrized approach, the $\alpha$ and $\beta$ indices denote the
planar indices.  For this reason, we don't expect the final results to
be too different for either approach.  For simplicity, we
choose to take the current operator to be the current between the $\alpha$th and
$\alpha+1$st planes for our derivations below, and we discuss how to get
the corresponding symmetrized results at the end.

The calculation of the local heat current operator is more complicated.
We adopt the same strategy as before though---first calculate the bulk
operator, and then extract a reasonable choice for the local operator.
To calculate the bulk heat current operator, we first need to determine
the energy polarization operator.  This is similar to the number polarization
operator, except it is weighted by the piece of the Hamiltonian associated
with the $\alpha i$ position.  This is easy to do for the interaction
and hybridization terms, the Coulomb potential energy terms, and the 
band offset terms, which are local, but is complicated for the 
hopping terms, which involve two lattice sites.  The procedure that is
used is to associate half of the hopping term between the two sites with the
local Hamiltonian at each of those two sites.  The energy polarization
term then becomes
\begin{eqnarray}
\Pi^E&=&\sum_\alpha\sum_{i\in\rm plane} \left [ \frac{1}{2}\sum_\beta
\sum_{j\in \rm plane}
\mathcal{H}_{{\rm hop}~\alpha i \beta j}+\mathcal{H}_{{\rm int}~\alpha i}
\right .\nonumber\\
&+&\left . 
\mathcal{H}_{{\rm charge}~\alpha i}+\mathcal{H}_{{\rm offset}~\alpha i}
\right ] {\bf R}_{\alpha i},
\label{eq: energy_pol}
\end{eqnarray}
where the hopping piece is divided into two as described above, and the
interaction piece includes all the local parts of the interacting 
Hamiltonian associated
with each lattice site. The bulk energy current operator is
$j^{\rm E}=i[\mathcal{H},\Pi^{\rm E}]$, and the heat-current operator is
$j^{\rm Q}=j^{\rm E}-\mu j$ because the heat is the energy measured relative
to the chemical potential.  The commutator is tedious to work out, but 
just involves straightforward algebra.  When it is finished, we have an 
expression for the bulk heat current, which can be organized into summations
that involve a plane $\alpha$ and the plane to the right (there is also a
hopping term involving operators at the $\alpha+2$ plane).  One
simply groups the terms together to find how to make an educated
guess for the local heat current operator.  The final results that we have 
are summarized below.  These are the proper local heat current operators
needed to satisfy the Jonson-Mahan theorem, as described below.  In all 
cases, we have $j^{\rm Q}=\sum_\alpha j^{\rm Q}_\alpha$.  Note that we
can form the symmetric version of the heat current operator as well, if desired,
but it is even more complex.

\begin{widetext}
For the Hubbard model, we have
\begin{eqnarray}
{\bf j}^{{\rm Q}}_\alpha&=&iat^\perp_{\alpha\alpha+1}\Biggr \{
-\sum_{ij\in \rm plane, \sigma}\frac{1}{2}
(t^\parallel_{\alpha ij}+t^\parallel_{\alpha+1 ij})
(c^\dagger_{\alpha+1 i \sigma}c^{}_{\alpha j \sigma}
-
c^\dagger_{\alpha i \sigma}c^{}_{\alpha +1 j \sigma})\nonumber\\
&-&
\frac{1}{2}t^\perp_{\alpha+1\alpha+2}\sum_{i\in \rm plane, \sigma}
(c^\dagger_{\alpha+2i \sigma}c^{}_{\alpha i \sigma}
-c^\dagger_{\alpha i \sigma}c^{}_{\alpha+2 i \sigma})
-
\frac{1}{2}t^\perp_{\alpha-1\alpha}\sum_{i\in \rm plane, \sigma}
(c^\dagger_{\alpha+1 i \sigma}c^{}_{\alpha-1 i \sigma}
-
c^\dagger_{\alpha-1 i \sigma}c^{}_{\alpha+1 i \sigma})
\nonumber\\
&+&\sum_{i\in\rm plane, \sigma}\left [
-\mu+\frac{1}{2}(V_\alpha+V_{\alpha+1})-
\frac{1}{2}(\Delta E_{F\alpha}+\Delta E_{F\alpha+1})\right ]
(c^\dagger_{\alpha+1 i \sigma}c^{}_{\alpha i \sigma}
-
c^\dagger_{\alpha i \sigma}c^{}_{\alpha+1 i \sigma})\nonumber\\
&+&
\frac{1}{2}\sum_{i\in \rm plane,  \sigma}
(U_\alpha c^\dagger_{\alpha i \bar \sigma}c_{\alpha i \bar \sigma}
+
U_{\alpha+1} c^\dagger_{\alpha+1 i \bar \sigma}  c_{\alpha+1 i \bar \sigma} )
(c^\dagger_{\alpha+1 i \sigma}c^{}_{\alpha i \sigma}
-
c^\dagger_{\alpha i \sigma}c^{}_{\alpha+1 i \sigma})\Biggr \} ,
\label{eq: local_heat_current_Hubbard}
\end{eqnarray}
where $\bar  \sigma=-\sigma$ denotes the spin state opposite to $ \sigma$.
For the Falicov-Kimball model, we find
\begin{eqnarray}
{\bf j}^{\rm Q}_\alpha&=&iat^\perp_{\alpha\alpha+1}\Biggr \{
-\sum_{ij\in \rm plane}\frac{1}{2}
(t^\parallel_{\alpha ij}+t^\parallel_{\alpha+1 ij})
(c^\dagger_{\alpha+1 i}c^{}_{\alpha j}-c^\dagger_{\alpha i}c^{}_{\alpha +1
j})
-\frac{1}{2}t^\perp_{\alpha+1\alpha+2}\sum_{i\in \rm plane}
(c^\dagger_{\alpha+2i}c^{}_{\alpha i}
-c^\dagger_{\alpha i}c^{}_{\alpha+2 i})\nonumber\\
&-&
\frac{1}{2}t^\perp_{\alpha-1\alpha}\sum_{i\in \rm plane}
(c^\dagger_{\alpha+1 i}c^{}_{\alpha-1 i}
-c^\dagger_{\alpha-1 i}c^{}_{\alpha+1 i})
+\frac{1}{2}\sum_{i\in \rm plane}
(U_\alpha w_{\alpha i}+U_{\alpha+1}w_{\alpha+1 i})
(c^\dagger_{\alpha+1 i}c^{}_{\alpha i}-c^\dagger_{\alpha i}c^{}_{\alpha+1 i})
\nonumber\\
&+&\sum_{i\in\rm plane}\left [ -\mu+\frac{1}{2}(V_\alpha+V_{\alpha+1})-
\frac{1}{2}(\Delta E_{F\alpha}+\Delta E_{F\alpha+1})\right ]
(c^\dagger_{\alpha+1 i}c^{}_{\alpha i}-c^\dagger_{\alpha i}
c^{}_{\alpha+1 i})\Biggr \} .
\label{eq: local_heat_current_text}
\end{eqnarray}
For the periodic Anderson model the commutation of the Hamiltonian with the
energy polarization operator gives
\begin{eqnarray}
{\bf j}^{{\rm Q}}_\alpha&=&iat^\perp_{\alpha\alpha+1}\Biggr \{
-\sum_{ij\in \rm plane, \sigma}\frac{1}{2}
(t^\parallel_{\alpha ij}+t^\parallel_{\alpha+1 ij})
(c^\dagger_{\alpha+1 i \sigma}c^{}_{\alpha j \sigma}
-
c^\dagger_{\alpha i \sigma}c^{}_{\alpha +1 j \sigma})\nonumber\\
&-&
\frac{1}{2}t^\perp_{\alpha+1\alpha+2}\sum_{i\in \rm plane, \sigma}
(c^\dagger_{\alpha+2i \sigma}c^{}_{\alpha i \sigma}
-c^\dagger_{\alpha i \sigma}c^{}_{\alpha+2 i \sigma})
-
\frac{1}{2}t^\perp_{\alpha-1\alpha}\sum_{i\in \rm plane, \sigma}
(c^\dagger_{\alpha+1 i \sigma}c^{}_{\alpha-1 i \sigma}
-
c^\dagger_{\alpha-1 i \sigma}c^{}_{\alpha+1 i \sigma})
\nonumber\\
&+&\sum_{i\in\rm plane, \sigma}\left [
-\mu+\frac{1}{2}(V_\alpha+V_{\alpha+1})-
\frac{1}{2}(\Delta E_{F\alpha}+\Delta E_{F\alpha+1})\right ]
(c^\dagger_{\alpha+1 i \sigma}c^{}_{\alpha i \sigma}
-
c^\dagger_{\alpha i \sigma}c^{}_{\alpha+1 i \sigma})\Biggr \}\nonumber\\
&+&
iaV^{hyb}_{\alpha}\frac{1}{2}\sum_{i\in \rm plane, \sigma}
\left[
t^\perp_{\alpha \alpha+1}
(f^\dagger_{\alpha i \sigma}c^{}_{\alpha+1 i \sigma}-c^\dagger_{\alpha+1 i
\sigma}f^{}_{\alpha i \sigma})
+
t^\perp_{\alpha \alpha-1}
(f^\dagger_{\alpha i \sigma}c^{}_{\alpha-1 i \sigma}-c^\dagger_{\alpha-1 i
\sigma}f^{}_{\alpha i \sigma}
)\right ]
\nonumber \\
&+&
iat^\perp_{\alpha\alpha+1}
\frac{1}{2}\sum_{i\in \rm plane,  \sigma}
\left[
(U_\alpha f^\dagger_{\alpha i \bar \sigma}f_{\alpha i \bar \sigma}
+
U_{\alpha+1} f^\dagger_{\alpha+1 i \bar \sigma}  f_{\alpha+1 i \bar \sigma} )
(f^\dagger_{\alpha+1 i \sigma}f^{}_{\alpha i \sigma}
-
f^\dagger_{\alpha i \sigma}f^{}_{\alpha+1 i \sigma})
\right ]
\label{eq: local_heat_current_PAM} .
\end{eqnarray}
\end{widetext}
The heat current operator depends on the
model being examined, because it involves commutators of the potential energy
with the energy polarization. We also subtract the chemical potential
multiplied by the number current from the energy current to get the heat
current.  One might have thought we should subtract the ``local chemical
potential''
multiplied by the local number current operator, but that would remove the
extra terms in the heat current arising from the electronic charge
reconstruction; one could have grouped those terms into either the Hamiltonian
or the local chemical potential---we chose the former, so we subtract only
$\mu j$.

Now we need to determine the $dc$ limit of the correlation functions $L_{ij}$
on the real axis.  The analytic-continuation procedure is identical to 
that for the bulk case. We start by defining a polarization
operator on the imaginary axis, then we analytically continue to the real axis,
we form the relevant transport coefficient, and then we take the limit of
the frequency going to zero. We denote four polarization operators
by $\bar L_{ij\alpha\beta}(i\nu_l)$ according to
\begin{eqnarray}
\bar L_{11\alpha\beta}(i\nu_l)&=&\int_0^\beta d\tau e^{i\nu_l\tau}\langle
\mathcal{T}_\tau j^{}_{\alpha}(\tau)j^{}_\beta(0)\rangle,
\nonumber\\
\bar L_{12\alpha\beta}(i\nu_l)&=&\int_0^\beta d\tau e^{i\nu_l\tau}\langle
\mathcal{T}_\tau j^{}_{\alpha}(\tau)j^{\rm Q}_\beta(0)\rangle,
\nonumber\\
\bar L_{21\alpha\beta}(i\nu_l)&=&\int_0^\beta d\tau e^{i\nu_l\tau}\langle
\mathcal{T}_\tau j^{\rm Q}_{\alpha}(\tau)j^{}_\beta(0)\rangle,
\nonumber\\
\bar L_{22\alpha\beta}(i\nu_l)&=&\int_0^\beta d\tau e^{i\nu_l\tau}\langle
\mathcal{T}_\tau j^{\rm Q}_{\alpha}(\tau)j^{\rm Q}_\beta(0)\rangle,
\label{eq: l_imag}
\end{eqnarray}
and the transport coefficients satisfy 
\begin{eqnarray}
L_{ij\alpha\beta}&=&
\lim_{\nu\rightarrow 0} \frac{1}{2i\nu}\left [ \bar L_{ij\alpha\beta}(\nu+i0^+)
-\bar L_{ij\alpha\beta}(\nu+i0^-)\right ]\nonumber\\
&=&\lim_{\nu\rightarrow 0} {\rm Re}[-i\bar L_{ij\alpha\beta}(\nu)/\nu]
\label{eq: lij_ac_form}
\end{eqnarray}
(the $ij$ subscripts here are 1 or 2, and not the planar site indices).
The generic notation $\mathcal{O}(\tau)=\exp[(\mathcal{H}-\mu\mathcal{N})\tau]
\mathcal{O}\exp[-(\mathcal{H}-\mu\mathcal{N})\tau]$ is used to indicate the
time dependence of the operators in Eq.~(\ref{eq: l_imag}).
The Jonson-Mahan theorem~\cite{jonson_mahan_1980,jonson_mahan_1990}
can be straightforwardly generalized to treat this case. Begin by defining
a generalized function
\begin{widetext}
\begin{eqnarray}
F_{\alpha\beta}(\tau_1,\tau_2,\tau_3,\tau_4)&=&\Bigr\langle\mathcal{T}_\tau
iat^\perp_{\alpha\alpha+1}\sum_{i\in \rm plane}\left [ 
c^\dagger_{\alpha+1i}(\tau_1)
c^{}_{\alpha i}(\tau_2)-c^\dagger_{\alpha i}(\tau_1)c^{}_{\alpha+1i}(\tau_2)
\right ]\nonumber\\
&\times&
iat^\perp_{\beta\beta+1}\sum_{j\in \rm plane}\left [c^\dagger_{\beta+1j}
(\tau_3)
c^{}_{\beta j}(\tau_4)-c^\dagger_{\beta j}(\tau_3)c^{}_{\beta+1j}(\tau_4)
\right ]  \Bigr\rangle.
\label{eq: jm_f}
\end{eqnarray}
Next, we determine the polarization operators by taking the appropriate
limits and derivatives.  Namely,
\begin{eqnarray}
\bar L_{11\alpha\beta}(i\nu_l)&=&\int_0^\beta d\tau_1 e^{i\nu_l\tau_1}
F_{\alpha\beta}(\tau_1,\tau_1^-,0,0^-),\nonumber\\
\bar L_{12\alpha\beta}(i\nu_l)&=&\int_0^\beta d\tau_1 e^{i\nu_l\tau_1} 
\frac{1}{2}
\left (\frac{\partial}{\partial \tau_3}-\frac{\partial}{\partial \tau_4}
\right ) F_{\alpha\beta}(\tau_1,\tau_1^-,\tau_3,\tau_4)
\Bigr |_{\tau_3=0,\tau_4=0^-}\nonumber\\
\bar L_{21\alpha\beta}(i\nu_l)&=&\int_0^\beta d\tau_1 e^{i\nu_l\tau_1} 
\frac{1}{2}
\left (\frac{\partial}{\partial \tau_1}-\frac{\partial}{\partial \tau_2}
\right ) F_{\alpha\beta}(\tau_1,\tau_2,0,0^-)
\Bigr |_{\tau_2=\tau_1^-}\nonumber\\
\bar L_{22\alpha\beta}(i\nu_l)&=&\int_0^\beta d\tau_1 e^{i\nu_l\tau_1} 
\frac{1}{4}
\left (\frac{\partial}{\partial \tau_1}-\frac{\partial}{\partial \tau_2}
\right )\left (\frac{\partial}{\partial \tau_3}-\frac{\partial}{\partial \tau_4}
\right )F_{\alpha\beta}(\tau_1,\tau_2,\tau_3,\tau_4)
\Bigr |_{\tau_2=\tau_1^-,\tau_3=0,\tau_4=0^-}.\nonumber\\
&~&\label{eq: jm_pol}
\end{eqnarray}
This result holds because the $(\partial_\tau-\partial_{\tau^\prime})/2$
operator converts the local number current operator into the local
heat current operator. To see this, we simply compute
\begin{eqnarray}
&~&\lim_{\tau^\prime\rightarrow\tau}
\frac{1}{2}\left ( \frac{\partial}{\partial\tau}-
\frac{\partial}{\partial\tau^\prime}\right ) 
iat^\perp_{\alpha\alpha+1}
\sum_{i\in{\rm plane}} \left [ c^\dagger_{\alpha+1 i}(\tau)c^{}_{\alpha i}
(\tau^\prime)-c^\dagger_{\alpha i}(\tau)c^{}_{\alpha+1 i} (\tau^\prime)\right ]
\nonumber\\
&~&=iat^\perp_{\alpha\alpha+1}\sum_{i\in{\rm plane}} \left \{
[\mathcal{H}-\mu\mathcal{N},c^\dagger_{\alpha+1 i}(\tau)]c^{}_{\alpha i}(\tau)
+
c^\dagger_{\alpha+1 i}(\tau)[\mathcal{H}-\mu\mathcal{N},c^{}_{\alpha i}(\tau)]
\right \},
\label{eq: jonson_mahan_op_ident}
\end{eqnarray}
\end{widetext}
which can be shown to be equal to $j^{\rm Q}_{\alpha}$ when the 
commutators are evaluated.  It is this critical identity that connects the 
local number and heat current operators that is a requirement for the
formalism to satisfy the Jonson-Mahan theorem.  The analytic continuation
is complex, because it involves four-time functions in the general case,
and a detailed proof of the Jonson-Mahan theorem appears in Appendix B.
Instead, we provide a direct constructive proof in DMFT here, where we
neglect the vertex corrections.  This is just a heuristic approach to
the full problem.

The first step is to evaluate the expectation values of the Fermionic
operators (in the definition of $F$) via contractions, because we neglect the
vertex corrections. This yields
\begin{eqnarray}
&~&F_{\alpha\beta}(\tau_1,\tau_2,\tau_3,\tau_4)=
a^2t_{\alpha\alpha+1}t_{\beta\beta+1}\nonumber\\
&~&\sum_{ij\in {\rm plane}}
\left \{ G_{\beta\alpha+1ji}(\tau_4-\tau_1)
G_{\alpha\beta+1ij}(\tau_2-\tau_3)\right.\nonumber\\
&-&G_{\beta+1\alpha+1ji}(\tau_4-\tau_1) G_{\alpha\beta ij}(\tau_2-\tau_3)
\nonumber\\
&-&G_{\beta\alpha ji}(\tau_4-\tau_1)G_{\alpha+1\beta+1ij}(\tau_2-\tau_3)
\nonumber\\
&+& \left.
G_{\beta+1\alpha ji}(\tau_4-\tau_1)G_{\alpha+1\beta ij}(\tau_2-\tau_3)\right\}.
\label{eq: f_wick}
\end{eqnarray}
Next, we need to determine a spectral representation for the off-diagonal
Green's function. Using the fact that
\begin{equation}
G_{\alpha\beta ij}(z)=-\frac{1}{\pi}\int d\omega
\frac{{\rm Im}G_{\alpha\beta ij}(\omega)}{z-\omega},
\label{eq: g_off_spec}
\end{equation}
with $z$ in the upper half plane (which can be shown by using the Lehmann
representation), says that
\begin{equation}
G_{\alpha\beta ij}(\tau)=-\frac{1}{\pi}\int d\omega T\sum_n
\frac{e^{-i\omega_n\tau}}{i\omega_n-\omega}{\rm Im}G_{\alpha\beta ij}(\omega).
\label{eq: g_off_tau_spec}
\end{equation}
Now we convert the sum over Matsubara frequencies into a contour integral
(that surrounds each Matsubara frequency, but does not cross the real axis---the
contour is then deformed into two contours, one running just above and the
other just below the real axis),
but we must be careful to ensure that the procedure is well-defined.
If $\tau<0$, then
\begin{eqnarray}
T\sum_n\frac{e^{-i\omega_n\tau}}{i\omega_n-\omega}&=&-\frac{i}{2\pi}
\int_C dz \frac{e^{-z\tau}}{z-\omega}f(z),\nonumber\\
&=&-\frac{i}{2\pi}\int_{-\infty}^{\infty}dz e^{-z\tau}f(z)\nonumber\\
&~&\left [
\frac{1}{z+i0^+-\omega}-\frac{1}{z-i0^+-\omega}\right ],\nonumber\\
&=&-e^{-\omega\tau}f(\omega).
\label{eq: tau_ident1}
\end{eqnarray}
This result is well-defined because the Fermi factor provides
convergence (asymptotically like $\exp[-\beta z]$) for $z\rightarrow\infty$
and the $\exp[-z\tau]$ term provides boundedness for $z\rightarrow -\infty$
when $\tau<0$.  Since $1-f(z)$ has the same poles as $f(z)$ on the imaginary
axis, with residues that have the opposite sign, and it behaves like
$\exp[\beta z]$ for $z\rightarrow -\infty$, one finds
\begin{equation}
T\sum_n\frac{e^{-i\omega_n\tau}}{i\omega_n-\omega}=e^{-\omega\tau}[1-f(\omega)],
\label{eq: tau_ident2}
\end{equation}
for $\tau>0$. The results in Eqs.~(\ref{eq: tau_ident1}) and
(\ref{eq: tau_ident2}) can then be substituted into
Eq.~(\ref{eq: g_off_tau_spec}) to get the final formula for the off-diagonal
Green's function
\begin{equation}
G_{\alpha\beta ij}(\tau)=\left \{ {\begin{array}{l l}
-\frac{1}{\pi}\int d\omega  {\rm Im}G_{\alpha\beta ij}(\omega)e^{-\omega\tau}
[1-f(\omega)],& \tau>0\\
-\frac{1}{\pi}\int d\omega  {\rm Im}G_{\alpha\beta ij}(\omega)e^{-\omega\tau}
[-f(\omega)],& \tau<0.
\end{array}}\right.
\label{eq: g_off_final}
\end{equation}

Now we note that we can restrict ourselves to the case
$\tau_1>\tau_2>\tau_3>\tau_4$ without loss of generality, because that
is the ordering needed to get the relevant correlation functions.
Then we employ Eq.~(\ref{eq: g_off_final}) in Eq.~(\ref{eq: f_wick})
and use the fact that the summations over the spatial indices for the planes can
be Fourier transformed, and then the momentum summation can be replaced by an
integration over the two-dimensional density of states, to yield
\begin{eqnarray}
F_{\alpha\beta}(\tau_1,\tau_2,\tau_3,\tau_4)&=&
\frac{a^2}{\pi^2}t^\perp_{\alpha\alpha+1}
t^\perp_{\beta\beta+1}\label{eq: f_tau}\\
&~&\int d\omega \int
d\omega^\prime\int d\epsilon^\parallel\rho^{2d}(\epsilon^\parallel)\nonumber\\
&\times&f(\omega)[1-f(\omega^\prime)]e^{-\omega(\tau_4-\tau_1)-\omega^\prime
(\tau_2-\tau_3)}\nonumber\\
&\times&\left \{ {\rm Im}G_{\beta\alpha}(\epsilon^\parallel,\omega)
{\rm Im}G_{\alpha+1\beta+1}(\epsilon^\parallel,\omega^\prime)\right. \nonumber\\
&~&+{\rm Im}G_{\beta+1\alpha+1}(\epsilon^\parallel,\omega)
{\rm Im}G_{\alpha\beta}(\epsilon^\parallel,\omega^\prime)\nonumber\\
&~&-{\rm Im}G_{\beta\alpha+1}(\epsilon^\parallel,\omega)
{\rm Im}G_{\alpha\beta+1}(\epsilon^\parallel,\omega^\prime)\nonumber\\
&~&\left. -{\rm Im}G_{\beta+1\alpha}(\epsilon^\parallel,\omega)
{\rm Im}G_{\alpha+1\beta}(\epsilon^\parallel,\omega^\prime)\right \}.
\nonumber        
\end{eqnarray}
Now we can evaluate the polarizations, and directly perform the analytic
continuation.  We Fourier transform the expression in Eq.~(\ref{eq: f_tau})
to get the Matsubara frequency representation.  Then we replace $i\nu_l$
by $\nu+i0^+$, we construct the transport coefficients on the
real axis, and we finally take the limit $\nu\rightarrow 0$ to
get the $dc$ response. The factor $(\partial_\tau-\partial_{\tau^\prime})/2$
gives a factor of $(\omega+\omega^\prime)/2$ which goes to $(\omega+\nu/2)$
after integrating over the delta function that arises in the analytic
continuation. Setting $\nu=0$ gives an extra power of $\omega$ in the
integrand for each derivative factor in the response coefficient. The end
result is
\begin{eqnarray}
L_{ij\alpha\beta}&=&\frac{a^2}{\pi}t^\perp_{\alpha\alpha+1}
t^\perp_{\beta\beta+1}
\int d\omega \left ( -\frac{d f(\omega)}{d\omega}\right ) \omega^{i+j-2}
\label{eq: l_coeff_final}\\
&~&\int d\epsilon^\parallel \rho^{2d}(\epsilon^\parallel)\nonumber\\
&\{ &{\rm Im}G_{\beta\alpha}(\epsilon^\parallel,\omega)
{\rm Im}G_{\alpha+1\beta+1}(\epsilon^\parallel,\omega)\nonumber\\
&+&{\rm Im}G_{\alpha\beta}(\epsilon^\parallel,\omega)
{\rm Im}G_{\beta+1\alpha+1}(\epsilon^\parallel,\omega)
\nonumber\\
&-&{\rm Im}G_{\beta\alpha+1}(\epsilon^\parallel,\omega)
{\rm Im}G_{\alpha\beta+1}(\epsilon^\parallel,\omega)\nonumber\\
&-&{\rm Im}G_{\alpha+1\beta}(\epsilon^\parallel,\omega)
{\rm Im}G_{\beta+1\alpha}(\epsilon^\parallel,\omega) \}.
\nonumber
\end{eqnarray}
This is the generalized Jonson-Mahan theorem for inhomogeneous systems described
by DMFT with vertex corrections neglected.
Note that the equality of $L_{12}$ with $L_{21}$ is the Onsager reciprocal
relation~\cite{onsager_1931}.

If one wants to work with symmetrized currents rather than the currents
between the $\alpha$ and $\alpha +1$st planes, then the Kubo formulas will
be changed slightly to take into account the symmetrized current operators.
These can be constructed directly from the correlation functions already
illustrated above, and it is a simple exercise to take care of the
relevant bookkeeping; we leave such details to the reader.

\section{Analysis of experiments}

With the expressions for the phenomenological coefficients that appear in
Eqs.~(\ref{eq: charge_phenom}) and
(\ref{eq: heat_phenom}) determined, we now can move onto evaluating the
transport in different cases of interest.  The first point that needs to be
emphasized is that the total number of electrons is always conserved in the
system, so the charge current is conserved, and cannot change from plane
to plane $\langle j^c_\alpha\rangle=\langle j^c_\beta\rangle$.  There is 
no such conservation law for the heat current though,
because the electrons can change the amount of heat that they carry
depending on their local environment. Hence, it is the boundary conditions that
we impose upon the heat current that determines how it behaves in a multilayered
nanostructure.  This point will become important as we analyze different
experimental situations.

\begin{figure}[th]
\centerline{\includegraphics[width=3.25in,angle=0]{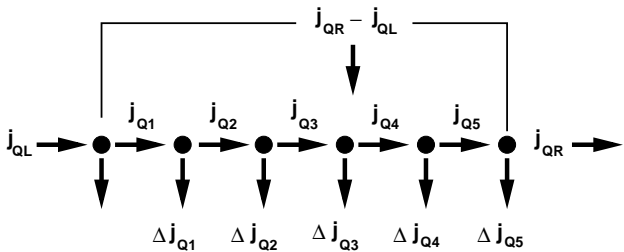}}
\vspace*{8pt}
\caption{Schematic diagram of the heat transfers in the Peltier effect.
The dots refer to different planes.  A heat current is incident from the
left.  As we go from one plane to another, the heat current changes,
as heat is transfered to or from a reservoir to maintain the
system at a constant temperature.  For example, we can examine the total heat
current transfered to the reservoirs ($j_{QR}-j_{QL}$), or we can examine the
average heat current that flows through the device $\sum_\alpha j_{Q\alpha}/N$,
where $N$ is the number of planes involved in the heat transfer.
\label{fig: peltier}}
\end{figure}

The first experiment we would like to analyze is the Peltier effect in
a multilayered nanostructure. We imagine that the nanostructure is attached
to a bath that maintains the entire structure at a fixed temperature,
and we then turn on an external electric field.  The Peltier effect is
the ratio of the heat current to the charge current.  A moment's reflection
will show that the heat current is not necessarily conserved in this system,
because we have to exchange heat with the reservoir to maintain a constant
temperature profile. Hence, it isn't even obvious what ratio should be
taken for the Peltier effect---the average heat current for the entire device
over the charge
current, the total change in the heat current over the charge current, or
the heat current transfered to the reservoir
over the charge current. We now show how to
determine all three of these results.

The starting point is the transport equations [(\ref{eq: charge_phenom})
and (\ref{eq: heat_phenom})] with $T_\alpha=T$ independent
of the plane number.  We first
determine the electric field by multiplying both sides of
Eq.~(\ref{eq: charge_phenom}) by the inverse $L_{11}$ matrix.  Since the
charge current is independent of the plane index, we find the electric field
satisfies
\begin{equation}
E_\alpha=\frac{1}{e^2a}\sum_\beta\left (L_{11}^{-1}\right )_{\alpha\beta}
\langle j^{\rm c}\rangle.
\label{eq: peltier1}
\end{equation}
Integrating the electric field over the $z$-coordinate, then
yields the voltage across the device, which allows us to extract the
resistance-unit-cell-area product via Ohm's law
\begin{equation}
R_na^2=\frac{1}{e^2}\sum_{\alpha\beta}\left ( L_{11}^{-1}\right )_{\alpha\beta}.
\label{eq: resistance}
\end{equation}
To find the heat current, we substitute the value of the electric field into 
Eq.~(\ref{eq: heat_phenom}), which yields
\begin{equation}
\langle j_\alpha^{\rm Q}\rangle
=-\frac{1}{|e|}\sum_{\beta\gamma}L_{21\alpha\beta}
\left (L_{11}^{-1}\right )_{\beta\gamma}\langle j^{\rm c}\rangle.
\label{eq: peltier2}
\end{equation}
This is all we need to analyze the Peltier effect of a nanostructure.
Note that the heat current generically will have $\alpha$ dependence, and hence
will vary from plane to plane (see Fig.~\ref{fig: peltier}).

The first question we can ask is how much heat is lost or gained by
the reservoir that is attached to the device to maintain isothermal conditions.
This is determined by the ratio of the difference in the heat current
at the right and the heat current at the left to the charge current.
In equations,
\begin{eqnarray}
\frac{\Delta \langle j^{\rm Q}\rangle}{\langle j^{\rm c}\rangle}&=&
\frac{\langle j_R^{\rm Q}\rangle -\langle j_L^{\rm Q}\rangle}{
\langle j^{\rm c}\rangle}\nonumber\\
&=&
-\frac{1}{|e|}\sum_{\beta\gamma}\left (L_{21R\beta}-L_{21L\beta}\right )
\left (L_{11}^{-1}\right )_{\beta\gamma}.
\label{eq: peltier3}
\end{eqnarray}
This would measure the net cooling or heating of the reservoir
by the device as the charge current flows.  Similarly, we could measure 
the average heat flow carried through the device
\begin{equation}
\frac{\langle j^{\rm Q}_{\rm ave}\rangle}{\langle j^{\rm c}\rangle}=
-\frac{1}{|e|}\frac{1}{N} \sum_{\alpha\beta\gamma} L_{21\alpha\beta}
\left (L_{11}^{-1}\right )_{\beta\gamma},
\label{eq: peltier4}
\end{equation}
where $N$ is the number of terms taken in the summation over the index $\alpha$.
This expression is analogous to the bulk Peltier effect, which measures
the ratio of the heat to charge current flows (which are independent
of position in a bulk system in linear response).

\begin{figure}[th]
\centerline{\includegraphics[width=3.25in,angle=0]{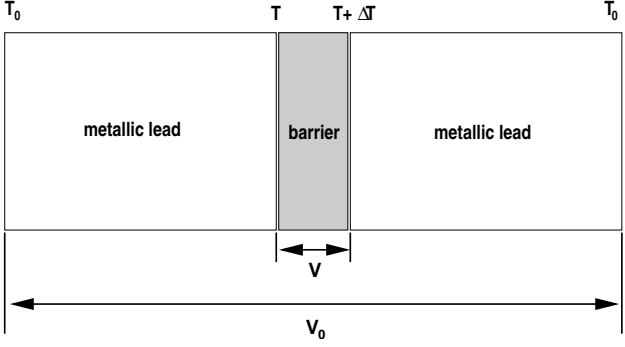}}
\vspace*{8pt}
\caption{Schematic diagram for how to measure the (relative) Seebeck effect.
The metallic leads are composed of the same material, and a voltage probe
is placed across the two ends which both are fixed at a temperature of $T_0$.
The voltage across those probes is $V_0$.  Since heat current will
flow across the voltmeter if both ends are not at the same temperature,
there is no way to directly measure the desired voltage $V$.  But, since the
change in voltage in the metallic lead in going from $T_0$ to $T$ on the
left hand side is exactly canceled by the change in voltage in going from
$T$ to $T_0$ on the right hand side, we find the difference between the
voltage $V$ and $V_0$ is just equal to the Seebeck coefficient of the
metallic lead multiplied by $\Delta T$.  Hence, since a measurement uses
$V_0$ instead of $V$, the Seebeck coefficient of the barrier is measured
{\it relative to} the Seebeck coefficient of the metallic lead.
\label{fig: seebeck}}
\end{figure}

Next we examine the Seebeck effect and a thermal conductivity experiment.
In both cases we work with an open circuit, so the total charge current
vanishes $\langle j^{\rm c}\rangle=0$. The Seebeck measurement is
subtle, because we
don't want to measure the voltage difference with probes at different
temperatures, because there will be a contribution from the $\nabla T d\mu/dT$
terms to the voltage drop (and there may be a thermal link allowing heat
to flow through the voltage probe).
An actual experiment uses thermocouple probes, where
one end of the probe is placed on the sample, and the other is placed in
a constant $T_0$ bath.
Two probes are needed to measure the voltage change and the
temperature at two points along the sample.  The net thermopower is measured
relative to the thermopower of the metal used in one of the legs of the
thermocouple (typically copper).  For details, see 
Refs.~\onlinecite{domenicali_1954}
and \onlinecite{nolas_etal}; a simpler schematic picture of this issue is
shown in Fig.~\ref{fig: seebeck}.  Alternatively, we can imagine the lead to the
left placed in a bath at temperature $T_0$, the interface plane on the left
held at temperature $T$, the interface on the right held at temperature
$T+\Delta T$, and the lead to the right held at temperature $T_0$.
The net effect on our analysis, if we assume the
thermopower of copper can be neglected (or of the ballistic lead in the
alternative picture), is that we neglect the $d\mu_\alpha/dT$
terms in our analysis (because the chemical potential at the probes
is at a constant temperature when the potential difference is measured).
With these caveats in mind, using Eq.~(\ref{eq: charge_phenom}), we find
\begin{equation}
E_\alpha=-\frac{1}{a|e|T}\sum_{\beta\gamma}
\left (L_{11}^{-1}\right )_{\alpha\beta}L_{12\beta\gamma}
(T_{\gamma+1}-T_\gamma).
\label{eq: seebeck1}
\end{equation}
Multiplying by $a$ and summing over $\alpha$ yields the voltage drop across
the device. We also need the temperature profile.  Substituting
Eq.~(\ref{eq: seebeck1}) into Eq.~(\ref{eq: heat_phenom}), and noting
that the heat current is conserved if the device is isolated and in the steady
state (implying heat cannot be transfered out of any plane, recall that the
Joule heating is a nonlinear effect) because the
system develops a temperature profile so that the heat current is conserved
through the device.  In this case, we can evaluate the temperature profile,
which satisfies
\begin{equation}
T_{\alpha+1}-T_\alpha=-T\sum_\beta\left ( M^{-1}\right )_{\alpha\beta}
\langle j^{\rm Q}\rangle,
\label{eq: seebeck2}
\end{equation}
with the matrix $M$ defined to be
\begin{equation}
M_{\alpha\beta}=L_{22\alpha\beta}-\sum_{\gamma\delta}L_{21\alpha\gamma}
\left ( L_{11}^{-1}\right )_{\gamma\delta} L_{12\delta\beta}.
\label{eq: seebeck3}
\end{equation}
Now we can sum Eq.~(\ref{eq: seebeck2}) over $\alpha$ to get the
temperature difference over the device.  Hence the Seebeck effect
becomes
\begin{equation}
\frac{\Delta V}{\Delta T}=-\frac{1}{|e|T}\frac{\sum_{\alpha\beta\gamma\delta}
(L_{11}^{-1})_{\alpha\beta}L_{12\beta\gamma}M^{-1}_{\gamma\delta}}
{\sum_{\alpha\beta}M^{-1}_{\alpha\beta}}.
\label{eq: seebeck4}
\end{equation}
Note that this is not equal to $1/T$ times the Peltier coefficient as in the
bulk [see Eqs.~(\ref{eq: peltier3}) and (\ref{eq: peltier4})].  
Instead, we have a weighting of the $L_{12}$ to $L_{11}$ ratio by the
matrix $M$, which is related to the thermal conductivity. This factor cancels
in the bulk, where the $M$ matrix depends on the difference of the
spatial coordinates, and the ${\bf q}=0$ response is independent of $M$
because the common factor in the Fourier transform will cancel out (as can
easily be proved by invoking the convolution theorem). If we do not measure
$\Delta V$ via thermocouples at constant $T$, then the $\Delta V$ term is
modified by a contribution from $d\mu_\alpha/dT$.  We do not discuss that
modification here, because it is not normally a technique used in 
measurements\cite{mahan_seebeck}.

The thermal conductance is evaluated in a similar way, but does not require
any subtlety in the measurement. We also work in an
open circuit, and the heat current is conserved, because we isolate the
system.  Now we measure the ratio of the heat current to the temperature
difference to find that the thermal conductance per unit area $K$ satisfies
\begin{equation}
K=-\frac{\langle j^{\rm Q}\rangle}{\Delta T}=\frac{1}{T\sum_{\alpha\beta}
(M^{-1})_{\alpha\beta}},
\label{eq: fourier1}
\end{equation}
and the thermal resistance-area product becomes
\begin{equation}
R_{\rm th}a^2=T\sum_{\alpha\beta}\left ( M^{-1}\right )_{\alpha\beta}.
\label{eq: fourier2}
\end{equation}.

Given all of the phenomenological parameters that enter into the transport
of a nanostructure, we are now in a position to be able to evaluate
things like the efficiency of a refrigerator, or of a power generator
(this requires evaluating heat flow while charge current is flowing, and
is more complex than the cases considered here).  The
final equations that result are quite complicated, and will not be shown here.
Note that it is necessary to perform such an exercise here, because in these
nanostructure devices, the thermoelectric cooling or power generation is
not determined solely by the bulk figure of merit of the constituent
pieces.  When quantum effects enter due to nanoscale structures, the 
situation is more complicated. But one can define an effective figure-of-merit
by constructing an effective Lorenz number from the ratio of the thermal to the
charge resistance and then evaluate an effective figure-of-merit from the 
Seebeck coefficient and the effective Lorenz number.  Even if this is done, 
it is not the same as the calculation of the efficiency of a real device, 
which simply is more complicated.

\section{Conclusions}

In this work we have shown how to derive the formalism for evaluating the
electronic contribution to the charge and thermal transport of a strongly
correlated nanostructure by determining the appropriate Kubo formulas for the
transport coefficients.  This analysis requires us to properly determine the
local current operators, which we do via a heuristic argument.  Using these
specific local current operators allows us to show that the Kubo formulas 
for the various heat transport coefficients are related by a generalization
of the Jonson-Mahan theorem to inhomogeneous systems.  We also describe how
nanostructures that will be used for heat transport must have an associated
electronic charge reconstruction, and we sketched how to solve for that
reconstruction using the DMFT approximation in the nanostructure.  We 
illustrated our results for models described by the Hubbard model, the
Falicov-Kimball model, and the periodic Anderson model, but they should hold
for any model that involves only local interactions.

It will be interesting to now solve some numerical problems and investigate 
the thermal transport coefficients for these systems.  Devices of particular
interest are those that have strongly correlated electrons, such as systems
with a Mott insulator, a doped Mott insulator, or Kondo metals in them.
Solving such problems is easiest to do for the Falicov-Kimball model because
it is the easiest model to solve numerically, but it should be feasible to
investigate the Hubbard model and the periodic Anderson model as well using
numerical renormalization group techniques, or quantum Monte Carlo plus
maximum entropy analytic continuations.  Such work will be presented
elsewhere.

\appendix

\section{Vanishing of currents for an equilibrium electronic charge
reconstruction}

We prove that the charge and heat current expectation
values vanish in equilibrium when there is an electronic charge reconstruction,
in the case where the self-energy is local.  The result should also hold for the
nonlocal case, by purely physical reasons, but we are not aware of a simple 
way to prove this result in the general case.

Because the multilayered system is translationally invariant in each plane, 
we can use a mixed basis, where we have a two-dimensional momentum describing
the planar degrees of freedom, and we use real-space to describe the
inhomogeneous $z$-direction\cite{potthoff_nolting_1999}.  Then, if 
we assume the self-energy is a local
function $\Sigma_\alpha$, that can vary from one plane to another, we have
the local Green's function satisfies\cite{freericks_2004}
\begin{eqnarray}
G_{\alpha\alpha}({\bf k}^\parallel,\omega)&=&1/\{
L_{\alpha}({\bf k}^\parallel,\omega)+R_{\alpha}({\bf k}^\parallel,\omega)
\label{eq: g_k_final}\\
&-&[\omega+\mu-V_\alpha+\Delta E_{\rm F \alpha}-\Sigma_\alpha(\omega)-
\epsilon_{\alpha{\bf k}^\parallel}^\parallel]\}
\nonumber
\end{eqnarray}
where $\epsilon_{\alpha{\bf k}^\parallel}^\parallel$ is the two-dimensional 
(planar) bandstructure on plane $\alpha$, and the left $L_\alpha$ and right
$R_\alpha$ functions are defined below.
The local Green's function on each plane is then
found by summing over the two-dimensional momenta, which can be replaced by an
integral over the two-dimensional density of states:
\begin{equation}
G_{\alpha\alpha}(\omega)=\int d\epsilon^{\parallel}_\alpha \rho^{2d}
(\epsilon^{\parallel}_\alpha) 
G_{\alpha\alpha}(\epsilon^{\parallel}_\alpha,\omega),
\label{eq: g_loc}
\end{equation}
The left function is defined to be
\begin{equation}
L_{\alpha-n}({\bf k}^\parallel,\omega)=
-\frac{G_{\alpha\alpha-n+1}({\bf k}^\parallel,\omega)
t^\perp_{\alpha-n+1\alpha-n}}
{G_{\alpha\alpha-n}({\bf k}^\parallel,\omega)}
\label{eq: ldef}
\end{equation}
and it satisfies the recurrence relation 
\begin{eqnarray}
L_{\alpha-n}({\bf k}^\parallel,\omega)&=&\omega+\mu-V_\alpha+
\Delta E_{\rm F \alpha}-\Sigma_{\alpha-n}(\omega)-
\epsilon_{\alpha-n{\bf k}^\parallel}^\parallel\nonumber\\
&-& \frac{t^\perp_{\alpha-n\alpha-n-1}t^\perp_{\alpha-n-1\alpha-n}}
{L_{\alpha-n-1}({\bf k}^\parallel,\omega)}.
\label{eq: l_recurrence}
\end{eqnarray}
We solve the recurrence relation by starting with the result for
$L_{-\infty}$, and then iterating Eq.~(\ref{eq: l_recurrence}).
In a similar fashion, we define a right function and a recurrence relation
to the right, with the right function satisfying
\begin{equation}
R_{\alpha+n}({\bf k}^\parallel,\omega)=
-\frac{G_{\alpha\alpha+n-1}({\bf k}^\parallel,\omega)t_{\alpha+n-1\alpha+n}}
{G_{\alpha\alpha+n}({\bf k}^\parallel,\omega)}
\label{eq: rdef}
\end{equation}
and the recurrence relation being
\begin{eqnarray}
R_{\alpha+n}({\bf k}^\parallel,\omega)&=&\omega+\mu-V_\alpha+
\Delta E_{\rm F \alpha}-\Sigma_{\alpha+n}(\omega)-
\epsilon_{\alpha+n{\bf k}^\parallel}^\parallel\nonumber\\
&-& \frac{t^\perp_{\alpha+n\alpha+n+1}t^\perp_{\alpha+n+1\alpha+n}}
{R_{\alpha+n+1}({\bf k}^\parallel,\omega)}.
\label{eq: r_recurrence}
\end{eqnarray}
We solve the right recurrence relation by starting with the result for
$R_{\infty}$, and then iterating Eq.~(\ref{eq: r_recurrence}).

In order to determine the current, we need to examine the off-diagonal,
nearest neighbor Green's functions, which satisfy
\begin{equation}
G_{\alpha+1\alpha}({\bf k}^\parallel,\omega)=
-\frac{G_{\alpha+1\alpha+1}({\bf k}^\parallel,\omega)
t^\perp_{\alpha+1\alpha}}{L_{\alpha}({\bf k}^\parallel,\omega)}\, ,
\end{equation}
and
\begin{equation}
G_{\alpha\alpha+1}({\bf k}^\parallel,\omega)=
-\frac{G_{\alpha\alpha}({\bf k}^\parallel,\omega)
t^\perp_{\alpha\alpha+1}}{R_{\alpha+1}({\bf k}^\parallel,\omega)}\, .
\end{equation}
Using the recursion relations for the Green's functions and for the $R$
and $L$ functions allows us to express the result for the Green's functions 
in terms
of $R_{\alpha+1}$ and $L_{\alpha}$.  Hence, we find
\begin{eqnarray}
G_{\alpha\alpha+1}({\bf k}^{\parallel},\omega)&=&G_{\alpha+1\alpha}
({\bf k}^{\parallel},\omega)\label{eq: zipper_current2}\\
&=&\frac{1}{L_{\alpha}({\bf k}^\parallel,\omega)
R_{\alpha+1}({\bf k}^\parallel,\omega)-
t^\perp_{\alpha\alpha+1}t^\perp_{\alpha+1\alpha}}.
\nonumber
\end{eqnarray}

Now, we are ready to show the expectation value of the number current operator
vanishes.  We can evaluate the expectation value of the number current
operator in Eq.~(\ref{eq: current_long}) by using the Green's functions
we have been describing above.  One finds
\begin{equation}
\langle j_\alpha\rangle=at^\perp_{\alpha\alpha+1}\int d\omega 
\sum_{\bf k^\parallel} [G_{\alpha+1\alpha}({\bf k}^\parallel,\omega)-
G_{\alpha\alpha+1}({\bf k}^\parallel,\omega)]=0,
\label{eq: expect_current_long}
\end{equation}
which vanishes because the two Green's functions are identical in value.  Note
that this vanishing of the current holds for arbitrary size electronic charge 
reconstruction, because it does not involve any linear-response assumption.
Similarly, since the heat-current operator is related to the number
current operator by a derivative with respect to time, and that derivative
analytically continues to an extra power of frequency in the integral over 
frequency, we also have that the heat-current operator expectation
value vanishes, because it involves adding an extra power of frequency
into the above integrals. This then completes the proof that the electronic
charge reconstruction does not have any currents flowing through it, so the
electric fields that enter the linear-response formalism are the external fields
only.

\section{Analytic continuation of the four-time response function needed
for the general Jonson-Mahan theorem proof}

We can restrict ourselves to the case
$\tau_1>\tau_2>\tau_3>\tau_4$ without loss of generality, because that
is the ordering needed to get the relevant correlation functions in 
Eqs.~(\ref{eq: jm_pol}). First of all, let us introduce a 
four-time correlation function (in real time) defined by
\begin{align}
   I_{ABCD}(t_1,t_2,t_3,t_4)&=I_{ABCD}(t_1-t,t_2-t,t_3-t,t_4-t)
\nonumber\\
      &=\langle A(t_1)B(t_2)C(t_3)D(t_4)\rangle,
\label{eq: I_ABCD_t}
\end{align}
where all operators are written in the Heisenberg representation 
$\mathcal{O}(t)=\exp[i(\mathcal{H}-\mu\mathcal{N})t]
\mathcal{O}\exp[-i(\mathcal{H}-\mu\mathcal{N})t]$ ($\hbar=1$). Its Fourier 
transform gives a four-time spectral density 
(with the constraint $\omega_1+\omega_2+\omega_3+\omega_4=0$ due to 
time-translation invariance of equilibrium correlation functions)
\begin{widetext}
\begin{align}
   I_{ABCD}(\omega_1,\omega_2,\omega_3,\omega_4)&=
   \int\limits_{-\infty}^{+\infty}\frac{dt_1}{2\pi}
   \int\limits_{-\infty}^{+\infty}\frac{dt_2}{2\pi}
   \int\limits_{-\infty}^{+\infty}\frac{dt_3}{2\pi}
   \exp[i\omega_1 t_1+i\omega_2 t_2+i\omega_3 t_3+i\omega_4 t_4]
I_{ABCD}(t_1,t_2,t_3,t_4)
\nonumber\\
   &=\frac1{\mathcal{Z}}\sum_{ilfp}e^{-\beta\varepsilon_i}
A_{il}B_{lf}C_{fp}D_{pi}
    \delta(\varepsilon_{il}+\omega_1)\delta(\varepsilon_{lf}+\omega_2)
\delta(\varepsilon_{fp}+\omega_3),
\label{eq: I_ABCD_w}
\end{align}
where $\varepsilon_i$ is the energy eigenvalue of the quantum many-body state 
$|i\rangle$
({\it i. e.}, the eigenvalue of the operator $\mathcal{H}-\mu\mathcal{N}$),
$\varepsilon_{il}$ satisfies $\varepsilon_{il}=\varepsilon_{i}-\varepsilon_{l}$,
$\mathcal{O}_{il}=\langle i|\mathcal{O}|l\rangle$ is the matrix element
of the operator $\mathcal{O}$ between the states $i$ and $l$, and 
$\mathcal{Z}=\sum_ie^{-\beta\varepsilon_i}$ is the partition function. 
The second line in Eq.~(\ref{eq: I_ABCD_w}) follows from the Lehmann
representation by inserting appropriate sets of complete states.
The spectral density satisfies the following cyclic permutation identities
\begin{align}
   I_{ABCD}(\omega_1,\omega_2,\omega_3,\omega_4)&=I_{BCDA}
(\omega_2,\omega_3,\omega_4,\omega_1)e^{\beta\omega_1}
   =I_{CDAB}(\omega_3,\omega_4,\omega_1,\omega_2)e^{\beta(\omega_1+\omega_2)}
   =I_{DABC}(\omega_4,\omega_1,\omega_2,\omega_3,)e^{-\beta\omega_4}
\label{eq: permut}
\end{align}
and transforms under Hermitian conjugation as
\begin{equation}
   I_{ABCD}(\omega_1,\omega_2,\omega_3,\omega_4)=
   [I_{D^\dag C^\dag B^\dag A^\dag }(-\omega_4,-\omega_3,-\omega_2,-\omega_1)
]^*,
\label{eq: Hermit}
\end{equation}
with the dagger indicating Hermitian conjugation of the associated operator.
Now, the generalized function in Eq.~(\ref{eq: jm_f}) can be defined for 
$\tau_1>\tau_2>\tau_3>\tau_4$ in terms of a generalized spectral density as
\begin{align}
   F_{\alpha\beta}(\tau_1>\tau_2>\tau_3>\tau_4)=
   \int\limits_{-\infty}^{+\infty}d\omega_1
   \int\limits_{-\infty}^{+\infty}d\omega_2
   \int\limits_{-\infty}^{+\infty}d\omega_3
   I_{\alpha\beta}(\omega_1,\omega_2,\omega_3,\omega_4)
   \exp[-\omega_1\tau_1-\omega_2\tau_2-\omega_3\tau_3-\omega_4\tau_4],
\label{eq: jm_ordrd}
\end{align}
where we introduce the total spectral density in terms of the partial 
one in Eq.~(\ref{eq: I_ABCD_w})
\begin{align}
   I_{\alpha\beta}(\omega_1,\omega_2,\omega_3,\omega_4)&=
-a^2t_{\alpha\alpha+1}^{\perp}t_{\beta\beta+1}^{\perp}
   \sum_{i\in \rm plane}\sum_{j\in \rm plane}
   \left[
   I_{c^\dagger_{\alpha+1i}c_{\alpha i}c^\dagger_{\beta+1j}c_{\beta j}}
(\omega_1,\omega_2,\omega_3,\omega_4)
   +I_{c^\dagger_{\alpha i}c_{\alpha+1i}c^\dagger_{\beta j}c_{\beta+1j}}
(\omega_1,\omega_2,\omega_3,\omega_4)
   \right. .
\nonumber\\
   &\left.{}-I_{c^\dagger_{\alpha+1i}c_{\alpha i}c^\dagger_{\beta j}
c_{\beta+1j}}(\omega_1,\omega_2,\omega_3,\omega_4)
   -I_{c^\dagger_{\alpha i}c_{\alpha+1i}c^\dagger_{\beta+1j}
c_{\beta j}}(\omega_1,\omega_2,\omega_3,\omega_4)
   \right] .
\label{eq: I_def}
\end{align}
Next, we can calculate the polarization operators in Eq.~(\ref{eq: jm_pol}) 
by taking the appropriate limits and derivatives.  Namely,
\begin{align}
   \bar L_{ij\alpha\beta}(i\nu_l)=
   \int\limits_{-\infty}^{+\infty}d\omega_1
   \int\limits_{-\infty}^{+\infty}d\omega_2
   \int\limits_{-\infty}^{+\infty}d\omega_3
   I_{\alpha\beta}(\omega_1,\omega_2,\omega_3,\omega_4)
   \frac{e^{-\beta(\omega_1+\omega_2)}-1}{i\nu_l+\omega_1-\omega_2}
   \left\{\begin{array}{ll}
   1                        & \text{for }ij=11\\
   \frac12(\omega_4-\omega_3)& \text{for }ij=12\\
   \frac12(\omega_2-\omega_1)& \text{for }ij=21\\
   \frac14(\omega_2-\omega_1)(\omega_4-\omega_3)& \text{for }ij=22\\
   \end{array}
   \right. .
\label{eq: L_fin_4}
\end{align}
Here, one notes that $\omega_4=-\omega_1-\omega_2-\omega_3$ due to the
constraint.
Then we replace $i\nu_l$ by $\nu+i0^+$ and construct the transport 
coefficients on the
real axis. Finally we take the limit $\nu\rightarrow 0$ to obtain the $dc$ 
response
\begin{align}
   L_{ij\alpha\beta}
   &=\lim_{\nu\rightarrow 0} \frac1{2i\nu}[\bar L_{ij\alpha\beta}(\nu+i0^+)-
\bar L_{ij\alpha\beta}(\nu-i0^+)]
   =\pi\beta\int\limits_{-\infty}^{+\infty}d\omega_2
\int\limits_{-\infty}^{+\infty}d\omega_4
   I_{\alpha\beta}(-\omega_2,\omega_2,-\omega_4,\omega_4)\omega_2^{i-1}
\omega_4^{j-1},
\label{eq: L_dc_4}
\end{align}
where, by using the identities in Eqs.~(\ref{eq: permut}) and 
(\ref{eq: Hermit}), the spectral density in 
Eq.~(\ref{eq: I_def}) can be reduced to the following expression
\begin{align}
   I_{\alpha\beta}(-\omega_2,\omega_2,-\omega_4,\omega_4)&=-2a^2
t_{\alpha\alpha+1}^{\perp}t_{\beta\beta+1}^{\perp}
   \sum_{i\in \rm plane}\sum_{j\in \rm plane}
   {\rm Re}\left[
   I_{c^\dagger_{\alpha+1i}c_{\alpha i}c^\dagger_{\beta+1j}c_{\beta j}}
(-\omega_2,\omega_2,-\omega_4,\omega_4)
   \right.
\nonumber\\
   &\left.
   {}-I_{c^\dagger_{\alpha+1i}c_{\alpha i}c^\dagger_{\beta j}
c_{\beta+1j}}(-\omega_2,\omega_2,-\omega_4,\omega_4)
   \right].
\label{eq: I_dc}
\end{align}
\end{widetext}
Eqs.~(\ref{eq: L_dc_4}) and (\ref{eq: I_dc})
are the generalization of the Jonson-Mahan theorem to nanostructures; the
integrands for the charge-charge, heat-charge, charge-heat, and heat-heat
current operator correlation functions are all related by powers of frequency.
One can also check that the Onsager reciprocal relation holds, where $L_{12}$ is
equal to $L_{21}$.  This follows by using the symmetry relations, and then
interchanging the dummy integration variables $\omega_2$ and $\omega_4$.

In the case where we neglect vertex corrections, the spectral densities in 
Eq.~(\ref{eq: I_dc}) are equal to
\begin{align}
   &I_{c^\dagger_{\alpha+1i}c_{\alpha i}c^\dagger_{\beta+1j}c_{\beta j}}
(-\omega_2,\omega_2,-\omega_4,\omega_4)
\nonumber\\
   &{}=I_{c_{\alpha i}c^\dagger_{\beta+1j}c_{\beta j}c^\dagger_{\alpha+1i}}
(\omega_2,-\omega_4,\omega_4,\omega_2,)
        e^{\beta\omega_4}
\nonumber\\
   &{}=\bar I_{c_{\beta j}c^\dagger_{\alpha+1i}}(\omega_4)
   \bar I_{c_{\alpha i}c^\dagger_{\beta+1j}}(\omega_4)e^{\beta\omega_4}
\delta(\omega_2-\omega_4)
\end{align}
and
\begin{align}
   &I_{c^\dagger_{\alpha+1i}c_{\alpha i}c^\dagger_{\beta j}c_{\beta+1j}}
(-\omega_2,\omega_2,-\omega_4,\omega_4)
\nonumber\\
   &{}=I_{c_{\alpha i}c^\dagger_{\beta j}c_{\beta+1j}c^\dagger_{\alpha+1i}}
(\omega_2,-\omega_4,\omega_4,-\omega_2)
        e^{\beta\omega_4}
\nonumber\\
   &{}=\bar I_{c_{\beta+1j}c^\dagger_{\alpha+1i}}(\omega_4)
   \bar I_{c_{\alpha i}c^\dagger_{\beta j}}(\omega_4)e^{\beta\omega_4}
\delta(\omega_2-\omega_4),
\end{align}
respectively, where
\begin{align}
   \bar I_{AB}(\omega)=-\frac{1}{\pi}f(\omega){\rm Im}G_{BA}(\omega)
\end{align}
are the single-particle spectral densities, and we obtain the same result as in 
Eq.~(\ref{eq: l_coeff_final}). In the general case, when vertex 
corrections are included, one can find the spectral densities in 
Eq.~(\ref{eq: I_ABCD_w}) from the 
multitime temperature (Matsubara) Green's function in Eq.~(\ref{eq: jm_f}) by 
employing spectral relations for the multitime correlation 
functions\cite{shvaika_ac}. The full derivation is complex and lengthy. In
the end it provides no new information with relation to the Jonson-Mahan
theorem, only an explicit formula for the charge conductivity matrix.  Hence,
we do not go through the derivation here.

\acknowledgments
We are grateful to P. Allen, G. Mahan, and L. Sham for useful discussions. 
Support from the N. S. F. under grant number DMR-0210717 is acknowledged.  
This publication is also based on work supported by Award No. UKP2-2697-LV-06 
of the U.S. Civilian Research \& Development Foundation (CRDF).


\begin{thebibliography}{99}
\bibitem{rontani_sham}
M. Rontani and L. J. Sham, Appl. Phys. Lett. {\bf 77}, 3033 (2001).
\bibitem{mahan_nano}
M. Bartkowiak and G. D. Mahan, in {\it Semiconductors and semimetals}
Vol. {\bf 71}, (Academic press, new York, 2001), p. 245.
\bibitem{jonson_mahan_1980}
M. Jonson and G. D. Mahan, Phys. Rev. B {\bf 21}, 4223 (1980).
\bibitem{jonson_mahan_1990}
M. Jonson and G. D. Mahan, Phys. Rev. B {\bf 42}, 9350 (1990).
\bibitem{hicks_dresselhaus_1993}
L. D. Hicks and M. S. Dresselhaus, Phys. Rev. B {\bf 47}, 12727 (1993).
\bibitem{ventkatasubramanian}
R. Ventkatasubramanian, E. Silvola, T. Colpitts, and
B. O'Quinn, Nature {\bf 413}, 597 (2001).
\bibitem{hubbard_1963}
J. Hubbard, Proc. R. Soc. (London) Ser. A {\bf 276}, 238 (1963).
\bibitem{falicov_kimball_1969}
L. M. Falicov and J. C. Kimball, Phys. Rev. Lett.  {\bf 22}, 997 (1969).
\bibitem{anderson_1961}
P. W. Anderson, Phys. Rev. {\bf  124}, 41 (1961).
\bibitem{schottky_1940}
W. Schottky, Phys. Z. {\bf 41}, 570 (1940).
\bibitem{freericks_sinis}
B. K. Nikoli\'c, J. K., Freericks and  P. Miller 
Phys. Rev. B {\bf 65}, 064529 (2002).
\bibitem{millis_okamoto}
S. Okamoto and A. J. Millis, Nature {\bf 428}, 630 (2004);
Phys. Rev. B {\bf 70}, 241104(R) (2004).
\bibitem{macdonald} 
W.-C. Lee and A. H. MacDonald, Phys. Rev. B {\bf 74}, 075106 (2006).
\bibitem{mahan_seebeck}
J. Cai and G. D. Mahan, Phys. Rev. B {\bf 74}, 075201 (2006).
\bibitem{potthoff_nolting_1999}
M. Potthoff and W. Nolting, Phys. Rev. B {\bf 59}, 2549 (1999).
\bibitem{ewald}
P. Ewald, Ann. Phys. (Leipzig) {\bf 64}, 253 (1921).
\bibitem{rpe}
E. N. Economou, {\it Green's functions in quantum physics}
(Springer-Verlag, Berlin, 1983).
\bibitem{freericks_2004}
J. K. Freericks, Appl. Phys. Lett. {\bf 84}, 1383 (2004);
Phys. Rev. B {\bf 70}, 195342 (2004).
\bibitem{chen_freericks}
L. Chen and J. K. Freericks, {\it unpublished}.
\bibitem{onsager_1931}
L. Onsager, Phys. Rev.  {\bf 37}, 405 (1931); Phys. Rev.  {\bf 38}, 2265 (1931).
\bibitem{fick_1855}
A. Fick, Poggendorff's Ann. Phys. u. Chemie {\bf 94}, 59 (1855).
\bibitem{einstein_1905}
A. Einstein, Ann. Phys. (Leipzig) {\bf 17}, 549 (1905).
\bibitem{nernst_1889}
W. H. Nernst, Z. Phys. Chem. {\bf 4}, 129 (1889).
\bibitem{smoluchowski_1906}
M. von Smoluchowski, Ann. Phys. (Leipzig) {\bf 21}, 756 (1906).
\bibitem{luttinger_1964}
J. M. Luttinger, Phys. Rev. {\bf 135}, A1505 (1964).
\bibitem{ohm_1827}
G. S. Ohm, {\it The Galvonic current investigated mathematically}
(J. G. F. Kniest\"adt, Berlin, 1827).
\bibitem{peltier_1834}
J. C. A. Peltier, Ann. Chim. Phys. {\bf 56}, 371 (1834).
\bibitem{fourier_1822}
J. B. J. Fourier, {\it Theorie analytique de la chaleur} 
(Firman, Didot, Paris, 1822);
English translation: {\it The analytic theory of heat}, translated by
A. Freeman (Cambridge University Press, Cambridge, 1878).
\bibitem{seebeck_1823}
T. J. Seebeck, Abhandlung der Deutschen Akademie der Wissenschaft zu Berlin,
265 (1823).
\bibitem{domenicali_1954}
C. A. Domenicali, Rev. Mod. Phys.  {\bf 26}, 237 (1954).
\bibitem{nolas_etal}
G. S. Nolas, J. Sharp and J. Goldsmid
{\it Thermoelectrics: Basic Principles and New Materials Developments}
(Springer-Verlag, Berlin, 2001).
\bibitem{shvaika_ac}
A. M. Shvaika, Condens. Matter Phys. \textbf{9}, XXX (2006) (in press);
{\it preprint arXiv:cond-mat/0604621}.



\end{thebibliography}
\end{document}